\documentclass[conference]{IEEEtran} 
\IEEEoverridecommandlockouts 

\usepackage{cite} 
\usepackage{url} 
\usepackage{textcomp} 
\usepackage{comment} 
\usepackage{balance} 
\usepackage{algorithmic} 
\usepackage[linesnumbered,ruled,vlined]{algorithm2e} 
\usepackage{amsmath,amssymb,amsfonts} 
\usepackage{mathtools} 
\usepackage{amsthm} 

\newtheorem{thm}{Theorem}
\newtheorem{lem}{Lemma}

\theoremstyle{remark}

\theoremstyle{definition}

\usepackage{graphicx} 
\usepackage{epstopdf} 
\usepackage{caption} 
\usepackage{subcaption} 
\usepackage{multirow} 
\usepackage{array}

\usepackage{enumitem} 
\usepackage{kantlipsum} 
\usepackage{setspace} 
\usepackage{xcolor} 
\usepackage{soul} 
\usepackage{booktabs} 
\definecolor{lhgreen}{RGB}{0,153,0}
\definecolor{lhred}{RGB}{204,0,0}
\definecolor{lhpurple}{RGB}{153,51,255}

\usepackage{xspace}
\def\name{\textit{Orchestra}\xspace}
\newcommand{\para}[1]{\noindent {\bf #1}} 

\begin{document}



 \title{Spatio-Temporal Parallelism for Diffusion Model Inference on Heterogeneous Multi-GPU Systems}

\author{\IEEEauthorblockN{
        Han Liang\IEEEauthorrefmark{2},
        Jiahui Zhou\IEEEauthorrefmark{2},
        Zicheng Zhou\IEEEauthorrefmark{2},
        Xiaoxi Zhang\IEEEauthorrefmark{2}\IEEEauthorrefmark{1},
        Xu Chen\IEEEauthorrefmark{2}\IEEEauthorrefmark{1}
        }
        \IEEEauthorblockA{
        \IEEEauthorrefmark{2}School of Computer Science and Engineering, Sun Yat-sen University, China
        }
        \IEEEauthorblockA{
        Email: 
        \{liangh68, zhoujh77, zhouzch7\}@mail2.sysu.edu.cn,
        \{zhangxx89, chenxu35\}@mail.sysu.edu.cn
        }
        \thanks{
        This work was supported in part by the National Natural Science Foundation of China (No. U25B2002, 62472460); Guangdong Basic and Applied Basic Research Foundation (No. 2023B1515120058, 2024A1515010161, 2023A1515012982); Guangzhou Basic and Applied Basic Research Program (No. 2024A04J6367); The Program for Guangdong Introducing Innovative and Entrepreneurial Teams (No. 2017ZT07X355).
        
        \IEEEauthorrefmark{1}Corresponding author: Xiaoxi Zhang, Xu Chen.
        }
}

\maketitle

\begin{abstract}
The widespread adoption of diffusion models for image generation necessitates efficient parallel inference to manage their substantial computational overhead. However, current parallel inference paradigms primarily target homogeneous clusters, often failing to maintain high efficiency in realistic, heterogeneous multi-GPU environments where hardware disparities and fluctuating background workloads cause severe straggler effects.
This paper introduces \name, a robust framework that orchestrates fine-grained parallelism across both temporal and spatial dimensions to harmonize computational loads in such settings. Temporally, \name employs a novel computation-aware step allocator using a tiered step reduction strategy, intelligently pruning denoising steps on slower devices after warmup phases and execution synchronization. Spatially, \name performs an elastic patch parallelism mechanism which adaptively adjusts the spatial workload intensity by assigning non-uniform image patches tailored to GPUs according to their computational capability.
Extensive experiments on load-imbalanced and heterogeneous clusters validate \name's efficacy in mitigating performance bottlenecks. Compared to patch parallelism, a state-of-the-art diffusion inference framework, our method reduces end-to-end latency by up to 45\% and significantly boosts resource utilization on heterogeneous GPUs. The source code is available at \url{https://github.com/liangh68/Orchestra}.
\end{abstract}




\section{Introduction}
\label{sec:intro}

Diffusion models, which were first proposed by~\cite{DDPM}, have emerged as a dominant force in generative artificial intelligence, sparking explosive growth across diverse applications. From generating photorealistic images~\cite{SD, SDXL, DiT} and videos~\cite{ho2022video,ho2022imagen,blattmann2023stable} to revolutionizing text-to-speech synthesis~\cite{popov2021grad,shen2023naturalspeech}, and multimodal content creation~\cite{MM-Diffusion,CoDi}, these models demonstrate unparalleled capabilities in synthesizing high-fidelity outputs from complex data distributions. 
However, as deployment shifts from research prototypes to real-world systems, the substantial computational demands and inherent iterative nature of diffusion sampling present a significant challenge: {\em prohibitive end-to-end inference latency}. This bottleneck not only hinders user experience in interactive applications but also limits the feasibility of deploying these powerful models on resource-constrained devices or in time-sensitive scenarios.



\begin{figure}[t]
    \centering
    \includegraphics[width=0.98\linewidth]{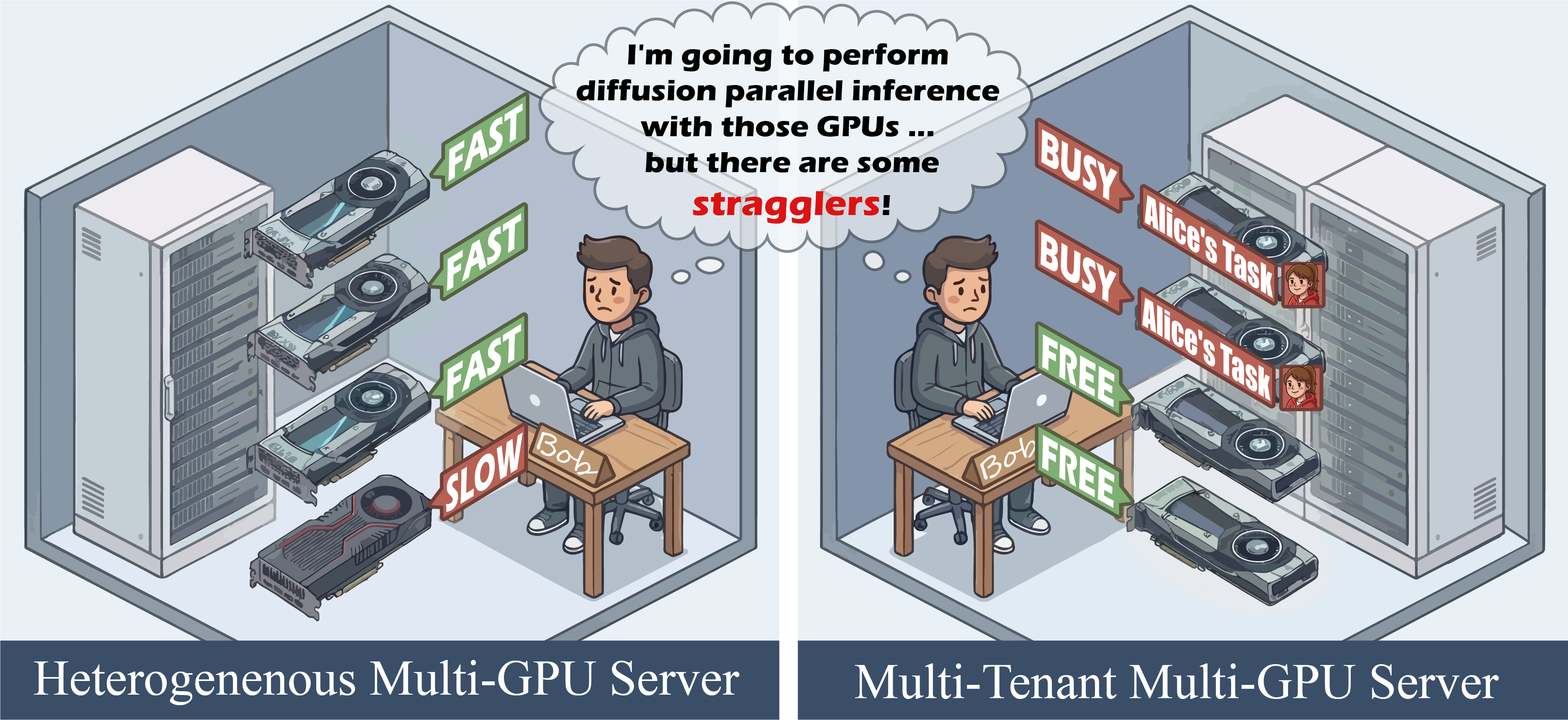}
    \caption{Bob wants to perform parallel inference for diffusion models on multi-GPU servers, but he encounters straggler issues caused by hardware heterogeneity (left) and multi-tenant resource contention (right). 
    }
    \vspace{-0.4cm}
    \label{fig:scene}
\end{figure}

Diffusion model inference (e.g., DDPM~\cite{DDPM}) is a denoising process consisting of a sequence of {\em steps} for each request, indexed as $T\to\cdots t+1, t\to\cdots 0$. In each step $t$, a trained neural network $\epsilon_{\theta}(\cdot,\cdot)$ predicts the noise 
based on the output (a noisy sample $\tilde{x}_{t+1}$) of step $t+1$ and the step $t+1$ itself. 
Then the diffusion scheduler will perform $\tilde{x}_{t} = \frac{\alpha_{t}}{\alpha_{t+1}} \tilde{x}_{t+1} - s(t,t-1) {\epsilon}_\theta(\tilde{x}_{t+1}, t+1)+\sigma_{t+1}\mathbf{z}$
to predict $\tilde{x}_t$, with $\tilde{x}_{t}$ defined as the {it estimated} noisy sample at timestep $t$ within the iterative reverse denoising trajectory.
This process progressively recovers the target data distribution from the pure noise.
The core bottleneck manifests as two interrelated factors: (1) the time per denoising step, and (2) the total number of required steps.
For the former, its duration is primarily constrained by the computational complexity of the model's forward propagation, which is often dominated by U-Net architectures~\cite{DDPM, SD, SDXL} or transformer blocks~\cite{DiT}. Strategies have centered on model compression techniques and GPU-level optimizations. 
Reducing the number of required sampling steps is equally critical. e.g., DDPM~\cite{DDPM} often necessitates thousands of steps for convergence. 

\begin{figure*}[t]
    \centering
    \begin{subfigure}[c]{0.30\textwidth}
        \centering
        \includegraphics[width=\linewidth]{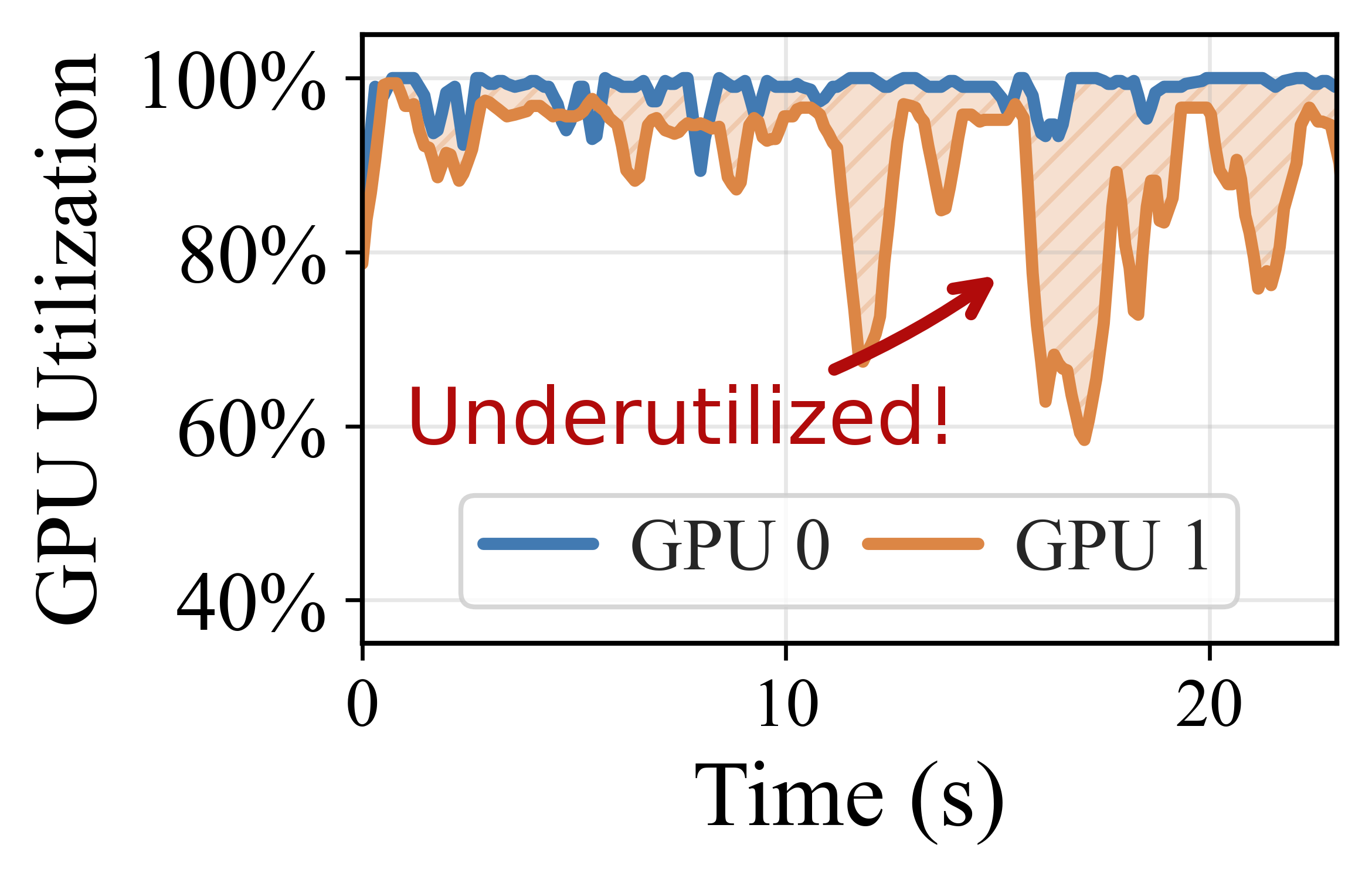}
        \caption{GPU utilization trace.}
        \label{fig:gpu_utils}
    \end{subfigure}
    \hfill 
    \begin{subfigure}[c]{0.30\textwidth}
        \centering
        \includegraphics[width=\linewidth]{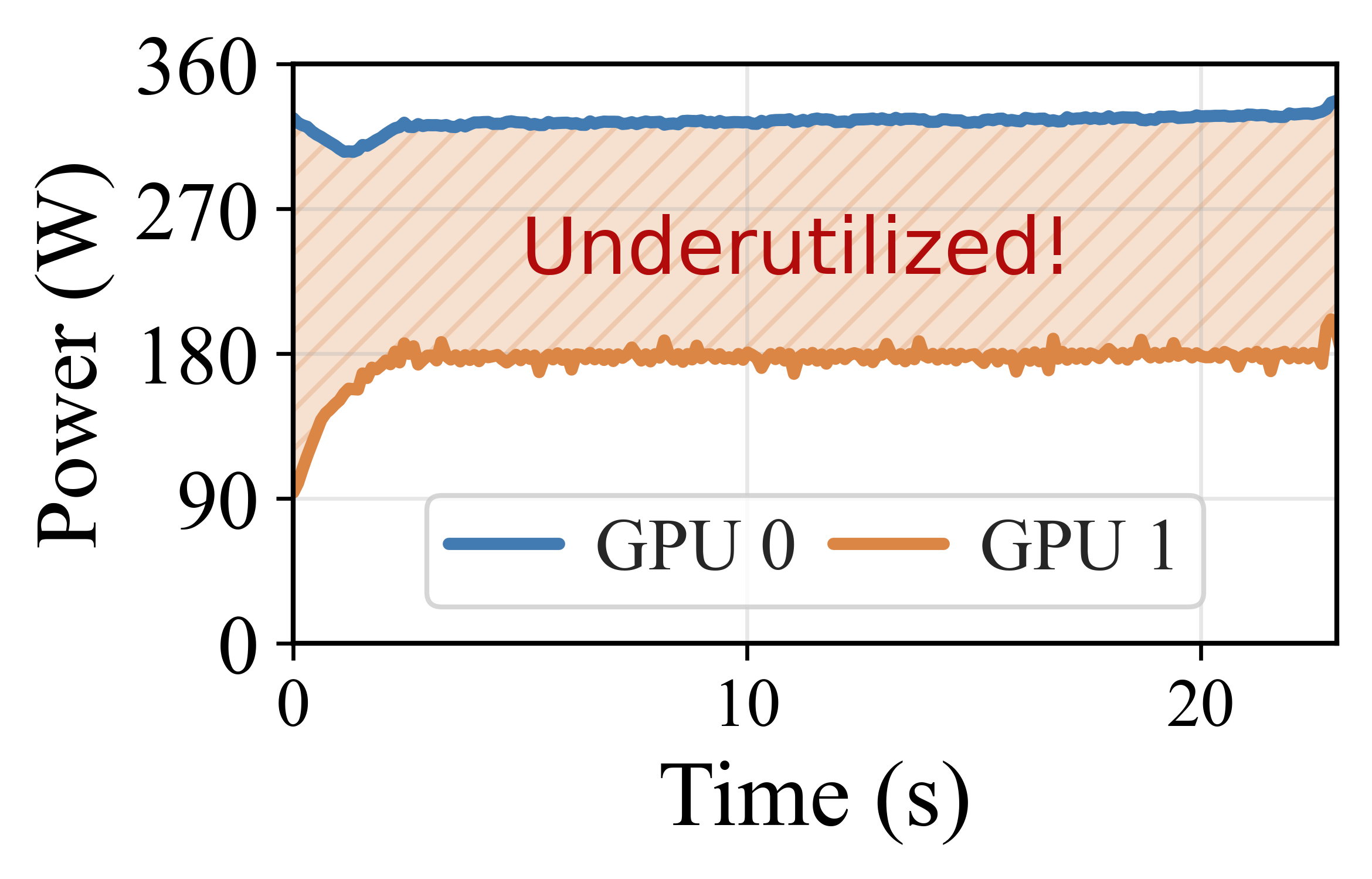}
        \caption{GPU power trace.}
        \label{fig:gpu_power}
    \end{subfigure}
    \hfill
    \begin{subfigure}[c]{0.38\textwidth}
        \centering
        \includegraphics[width=\linewidth]{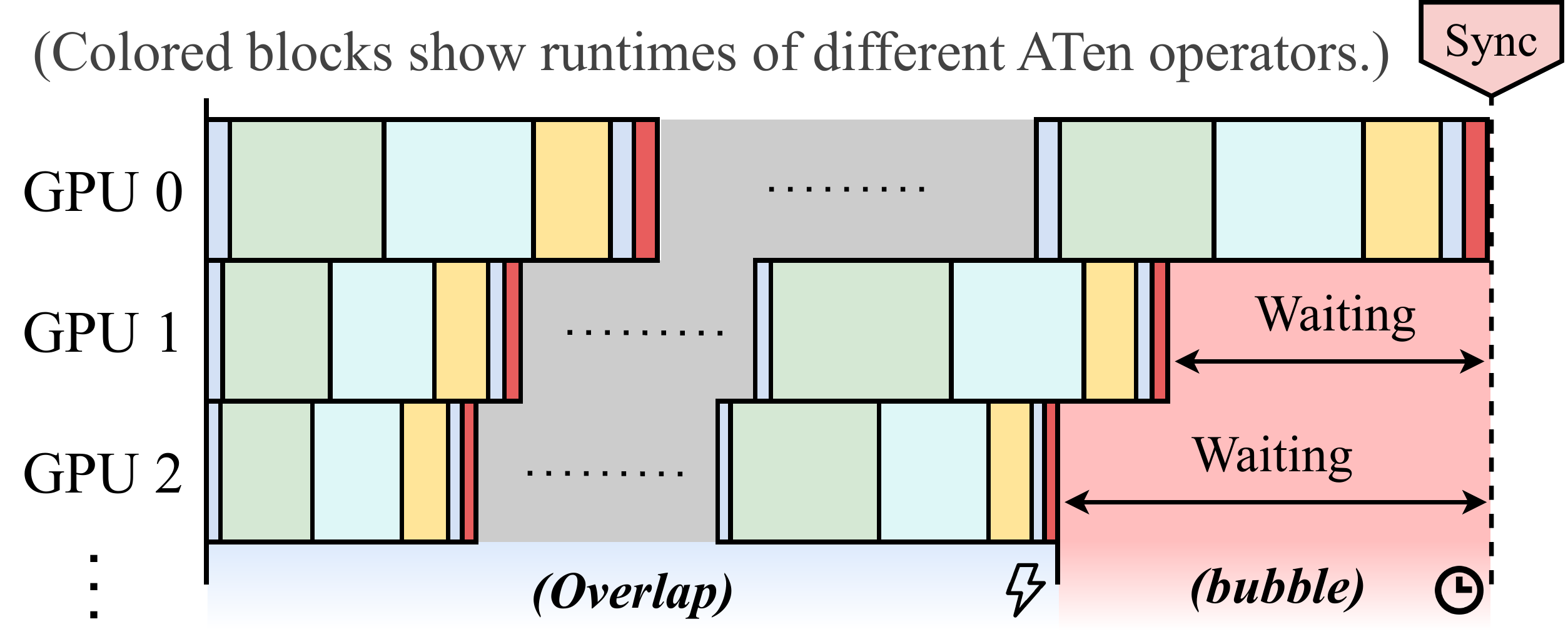}
        \caption{Workflow illustration.}
        \label{fig:motivation1}
    \end{subfigure}

    \caption{Performance degradation due to load imbalance. (a) \& (b): Heterogeneous background loads across GPUs lead to significant disparities in utilization and power consumption, when using homogeneous patch partitioning. (c) Blocking bottlenecks demonstrate the performance loss incurred when faster GPUs must wait for the bottlenecked GPU to complete.}
    \label{fig:combined_motivation}
\end{figure*}

\para{Advanced parallelism for diffusion inference.} To fundamentally decouple sampling latency from iterative step constraints, multi-GPU parallel inference paradigms~\cite{distrifusion, pipefusion, asyncdiff, parastep, xDiT, shih2023parallel} have emerged as a transformative approach.
Key innovations in this area include: DistriFusion~\cite{distrifusion}, which proposes {\em patch parallelism} with asynchronous stale activation reuse (stored in a buffer on each GPU) to achieve near-linear speedup; PipeFusion~\cite{pipefusion}, deploying a chunked pipeline strategy that slices transformer layers and then also leverages patch parallelism to optimize memory scaling for high-resolution DiT-based models; and AsyncDiff~\cite{asyncdiff}, which divides the model into several blocks with roughly equal computational complexity and assigns them to multiple GPUs for pipeline parallel execution. 
Although these methods achieve near-linear speedups in homogeneous clusters with dedicated resources, their efficacy sharply degrades in systems with heterogeneous GPU speeds. As depicted in Figure~\ref{fig:gpu_utils}, such heterogeneity, caused by hardware differences or imbalanced task occupancies across GPUs, increases inference latency and undermines resource utilization. 
Figure~\ref{fig:motivation1} further shows how this computational imbalance across devices cascades into blocking bottlenecks via inevitable per-step synchronization. These limitations underscore the necessity of adaptive frameworks robust to heterogeneity in both hardware and workloads.




\para{Existing multi-GPU methods for non-diffusion inference} fall into two categories. The first focuses on system-level scheduling to maximize throughput across multiple tasks~\cite{choi2022serving,tang2021aeml,zhang2024hap,zhang2024mixtran}, without accelerating individual requests. The second targets single-request acceleration using pipeline~\cite{mohammed2020distributed,hu2022distributed,kong2025ppipe}, tensor~\cite{shoeybi2019megatron}, or sequence parallelism~\cite{jacobs2023deepspeed,fang2024usp}. While effective for transformer inference, these approaches exploit only spatial parallelism and overlook temporal dynamics, an essential characteristic of diffusion model inference.


Therefore, we introduce Spatio-Temporal Adaptive Diffusion Inference, \name, the first diffusion-tailored inference framework that directly leverages iterative steps, the fundamental building blocks of each inference request, to mitigate GPU idle periods arising from system heterogeneity, complemented by patch parallelism as a supplementary strategy. 
Our key insight is that: Temporal adaptation is too coarse for precise balancing, while spatial adaptation alone triggers excessive kernel overhead on lagging devices; only their synergy enables fine-grained alignment without creating waiting bubbles.
This dual-dimensional method effectively minimizes inference latency while preserving image generation quality. To achieve this, \name makes the following technical contributions. 

\begin{itemize}
    \item \textbf{First heterogeneity-aware distributed diffusion model inference system}: To the best of our knowledge, this is the first work to study how state-of-the-art diffusion models can be distributed and accelerated under {\em system heterogeneity}. Using SDXL\cite{SDXL}, one of the most representative diffusion models,  we implement a distributed image generation system with adaptive patch parallelism and step-specific model synchronization to efficiently exploit multi-GPU platforms with diverse computational capabilities across devices. Experiments demonstrate that \name reduces inference latency by 27.01\% -- 45.36\%.
    \item \textbf{First fine-grained step and patch parallelism strategy.} 
    We are the first to propose a spatio-temporal adaptive parallelism mechanism optimizing both step and patch size across devices. 1) Temporally, \name decouples denoising trajectories across devices by assigning different denoising steps after shared warm-up phases, according to profiled GPU speeds and a tiered step reduction strategy. 
    It preserves fidelity while reducing desynchronization and idle time. 
    2) Spatially, it complementarily employs elastic patch size mending, scaling partitioned sub-regions to further balance compute load. This dual-axes adaptation fundamentally replaces static patch parallelism with a more flexible parallel approach, ensuring state consistency across uneven devices.
    \item \textbf{Theoretical Guarantee}: To facilitate analysis, we investigate DDIM~\cite{DDIM}, an accelerated diffusion inference framework that reduces DDPM's inference steps by a pre-determined factor (detailed in Section~\ref{ssec:ddim}). We derive a theoretical asymptotic bound of latent difference across consecutive denoising steps, which is not provided in the original DDIM work. Building on these results, we show that \name maintains the same order of asymptotic bound as DistriFusion. While DistriFusion operates in homogeneous multi-GPU systems, \name can serve heterogeneous multi-GPU environments.

\end{itemize}

\section{Background and Related Work}
\label{sec:background}

\subsection{Diffusion Model and Inference Acceleration}
\label{ssec:ddim}
Standard diffusion model inference is centered on a gradual denoising process to synthesize data~\cite{DDPM}. 
As we elaborate in Section~\ref{sec:intro},
it starts from a Gaussian noise $\tilde{x}_T$ and iteratively takes each $\tilde{x}_t$ as input of the neural network to generate $\tilde{x}_{t-1}$, and so on, a.k.a. denoising back to the target distribution over a number of steps. 

\begin{figure}[t]
    \centering
    \includegraphics[width=0.98\linewidth]{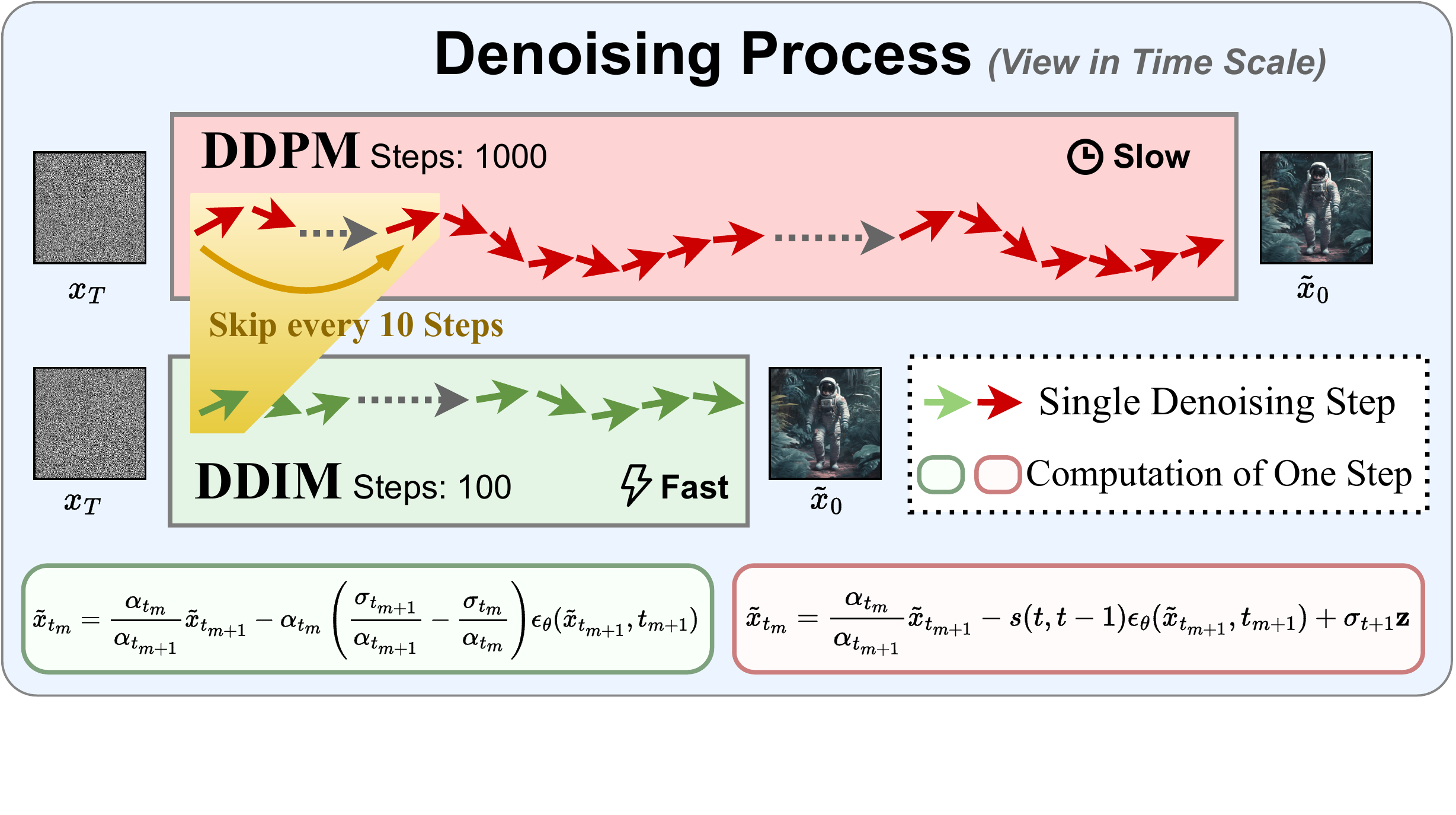}
    \vspace{-0.6cm}
    \caption{Architectural comparison between Denoising Diffusion Probabilistic Models (DDPM) and Denoising Diffusion Implicit Models (DDIM).
    }
    \vspace{-0.3cm}
    \label{fig:DDPMvsDDIM}
\end{figure}

\para{Diffusion inference acceleration.} Latent diffusion models~\cite{SD} improve efficiency by denoising in a compressed latent space, enabling prompt-guided high-resolution generation. While the complexity of diffusion is reduced from the pixel space of the original image to the latent space, the computational overhead of executing thousands of steps still remains a prohibitive problem. To mitigate this, many sampling acceleration strategies have emerged: DDIM~\cite{DDIM} reformulates the reverse process as deterministic ordinary differential equations (ODEs)\cite{kingma2021variational, DPMsolver}, enabling significant step reduction while preserving image quality. Lu et al.~\cite{DPMsolver, DPMsolverplus} identified the semi-linear structure of diffusion ODEs and devised higher-order solvers. Moreover, while significant research focuses on optimizing solvers for diffusion SDEs~\cite{DDPM, song2020score, xue2023sa, DPMsolverplus, bao2022analytic}, ODE-based solvers remain the preferred choice.
As illustrated in Figure~\ref{fig:DDPMvsDDIM}, the first-order solver~\cite{DPMsolver} proposed the following definition showing that $T$ steps corresponding to those in DDPM can be grouped into $M$ blocks, each of which consists of $\frac{T}{M}$ steps, and then the predicted noisy sample $\tilde{x}_{t_m}$ can be computed as follows
\begin{lem}[DPM-Solver-1~\cite{DPMsolver}, DDIM]
\label{lem:DDIM}
    Given $x_T$ and $M+1$ steps $\{t_m\}^{M}_{m=0}$ decreasing from $t_0=T$ to $t_M=0$. Then the sequence $\{\tilde{x}_{t_m}\}^{M}_{m=1}$ is computed as follows:
    \begin{align}
    \label{eq:ddimupdate}
    \tilde{{x}}_{t_m}&=\frac{\alpha_{t_{m}}}{\alpha_{t_{m-1}}}\tilde{{x}}_{t_{m-1}}-\sigma_{t_{m}}(e^{h_{m}}-1){\epsilon}_\theta(\tilde{{x}}_{t_{m-1}},t_{m-1}), 
    \end{align}
    where $h_m=\lambda_{t_m}-\lambda_{t_{m-1}}$ and $\lambda_{t_{m}}=\log(\alpha_{t_{m}}/\sigma_{t_{m}})$. 
\end{lem}

\para{Insight.} This line of work on reducing inference steps offers us the following insight: during multi-GPU parallel inference, we can leverage diverse step counts to tackle the heterogeneity of GPUs. 
But the challenge is non-trivial, as the inference quality may deteriorate with various denoising steps across distributed inference nodes due to inference results hard to align with each other.

\subsection{Diffusion Inference Frameworks on Multi-GPU Systems}

Existing multi-GPU parallelism algorithms for diffusion models primarily focus on spatial partitioning or pipeline optimizations. DistriFusion~\cite{distrifusion} is an advanced diffusion inference framework that partitions an input noise into uniform patches distributed across GPUs, leveraging patch parallelism to reuse stale activations from the previous inference step for asynchronous communication. However, its uniform patching strategy assumes power-of-two GPU numbers and homogeneous GPU capabilities, causing load imbalance when devices exhibit computational disparities. PipeFusion~\cite{pipefusion} extends this to DiTs~\cite{DiT} via patch-level pipeline parallelism, distributing DiT blocks across devices and reusing stale features to minimize communication. Although memory-efficient for large DiTs, its rigid pipeline stages lack adaptability in computational heterogeneity scenarios and rely on static pre-segmentation. Moreover, load imbalance occurs in PipeFusion when the number of transformer layers is not a multiple of the number of GPUs. Similarly, AsyncDiff~\cite{asyncdiff} pre-divides model layers for pipelined execution, while ParaStep's~\cite{parastep} reuse-then-predict mechanism reduces communication overheads.

Unfortunately, as demonstrated in Figure~\ref{fig:combined_motivation}, these frameworks suffer from straggler effects under system heterogeneity. 
We take a step forward by designing a novel temporal adaptation strategy to mitigate the stragglers' GPU idle time.

\subsection{Multi-GPU System for Non-Diffusion DL Inference}
\label{subsec:jiahuipart}

The escalating computational demands of deep generative models have driven the widespread adoption of multi-GPU systems to accelerate inference\cite{zhong2024poster,huang2024alleviating}. Existing efforts follow two main directions: multi-task throughput maximization~\cite{choi2022serving,tang2021aeml,zhang2024hap,zhang2024mixtran}, and single-request latency reduction. The first direction focuses on maximizing cluster-wide throughput for concurrent requests. Early systems for homogeneous clusters, such as Clipper~\cite{Clipper} and Nexus~\cite{Nexus}, employed request batching and multi-tenant scheduling. AEML~\cite{tang2021aeml} targets heterogeneous environments and dynamically redistributes workloads based on real-time performance profiling.


This work focuses on accelerating a single inference task, for which prior work explores several parallelism strategies.
(1) \textit{Pipeline parallelism} splits models into sequential stages. Systems like PPipe~\cite{kong2025ppipe} adapt this idea to heterogeneous GPUs by balancing stage sizes and batch allocations; similar approaches extend to DAG-structured models on edge devices~\cite{mohammed2020distributed,hu2022distributed}. 
PASTA\cite{wupasta} extends this technique to vertical federated learning with staleness control, overlapping communication and computation to accelerate training.
(2) \textit{Tensor parallelism}, introduced by Megatron-LM~\cite{shoeybi2019megatron}, partitions weight matrices within layers to enable parallel computation. Extensions like HeteGen~\cite{zhao2024hetegen} adapt this for CPU–GPU hybrid setups to mask memory bottlenecks.
(3) \textit{Sequence parallelism} partitions the input data (i.e., the token sequence) across devices. This has become a key technique for handling long contexts in LLMs. State-of-the-art approaches like DeepSpeed-Ulysses~\cite{jacobs2023deepspeed} use efficient all-to-all collectives, while others like Ring-Attention~\cite{fang2024usp} use P2P communication to compute global attention on partitioned data. 
Despite their strengths, these approaches primarily exploit spatial parallelism within individual inference steps, as they do not target diffusion models.  
Our work directly leverages denoising steps, the core component of diffusion inference, and we introduce a novel spatio-temporal adaptation parallelism strategy accordingly, enabling adaptive step reduction and patch resizing across heterogeneous GPUs.

\section{Method}
\label{sec:method}
\subsection{Overview}
\label{subsec:overview}

Existing parallel diffusion inference schemes often perform poorly on heterogeneous or partially utilized multi-GPU systems, exhibiting significant latency overheads that we later detail in our evaluation (see Sec.~\ref{sec:experiment}).
Our analysis in Fig.~\ref{fig:combined_motivation} reveals that static partitioning strategies, while effective for homogeneous systems, fail to accommodate hardware disparity, leading to severe GPU underutilization and performance bottlenecks.
Therefore, realizing efficient parallelization demands flexible and fine-grained adjustments to each GPU’s temporal computation depth (steps) and spatial computation scope (patch size). 

Inspired by DDIM and patch parallelism, we propose {\em Spatio-Temporal Adaptive Diffusion Inference (\name)}, a novel inference acceleration algorithm tailored for diffusion models deployed on heterogeneous multi-GPU systems. The core innovation of \name lies in its dual-axis adaptive scheduling mechanism: (1) {\em Temporal Adaptation}, which adaptively determines the number of steps $M_i$ assigned to each GPU $i$, and (2) {\em Spatial Adaptation}, which adaptively allocates a patch $p_i$ of size $P_i$ to each GPU $i$ as a fine-grained complement to the temporal step partitioning.

\begin{figure*}[t] 
    \centering 
    \includegraphics[width=0.98\textwidth]{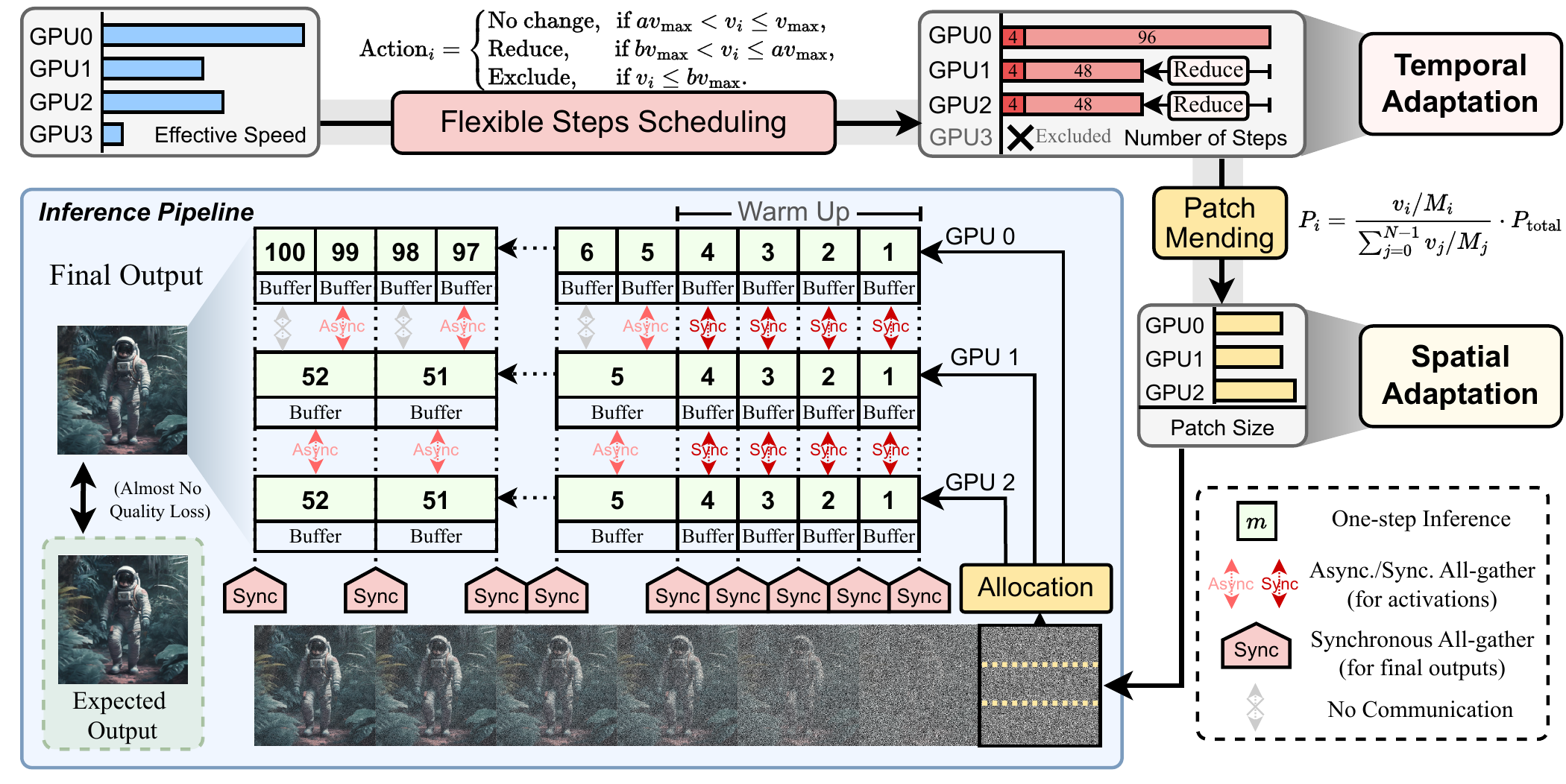} 
    \caption{
        Overview of \name.
    } 
    \label{fig:overview} %
\end{figure*}

\para{Line-by-line Algorithm Interpretation.} The pseudocode in Algorithm~\ref{alg:ours} outlines \name. It begins with temporal adaptation by computing distinct steps for each GPU, prioritizing larger step allocations to faster devices to balance processing time (lines 1-3). Subsequently, spatial adaptation determines patch sizes based on each GPU’s effective speed and allocated steps to further equalize the workload (lines 4-6). During inference, all GPUs first conduct warm-up steps where both buffer and output are gathered synchronously (lines 9-12). The inference loop then operates adaptively: as shown in Figure~\ref{fig:overview}, for faster GPUs, the system alternates between synchronizing to obtain noisy outputs and reusing the noisy output from the previous step without synchronization (lines 13-21); meanwhile, slower GPUs consistently perform synchronization after each local computation (lines 22-24). The final denoised image is returned after this process (line 25).
By evaluating both the inherent computational heterogeneity and the current load state across the system prior to inference, \name intelligently allocates computational resources in a load-balanced manner. Figure~\ref{fig:overview} illustrates the high-level architecture of the \name system, which coordinates heterogeneous GPUs and performs spatio-temporal resource allocation.

\begin{algorithm}[t]
    \caption{\name: Spatio-Temporal Adaptive Diffusion Inference}
    \label{alg:ours}
    \SetAlgoVlined
    \SetKwInOut{Input}{Input}
    \SetKwInOut{Output}{Output}
    \Input{An initial noise $x_{t_0}$, a pre-trained diffusion model $\epsilon_\theta(\cdot,\cdot)$, effective speed of all GPUs $\{v_i\}_{i=0}^{N-1}$, a base step number $M_{\text{base}}$, and a warm-up step number $M_{\text{warmup}}$}
    \Output{The final image $\tilde{x}_{t_M}$ predicted by \name.}
    // Step 1: Temporal Adaptation\\
    \For{GPU $i=0$ to $N-1$}{
        Get steps $M_i$ by Equation~\eqref{eq:steps_coordinating}.
    }
    // Step 2: Spatial Adaptation\\
    \For{GPU $i=0$ to $N-1$}{
        Get patch size by $P_i = \frac{v_i/M_i}{\sum_{j=0}^{N-1} v_j/M_j} \cdot P_{\text{total}}$
    }
    // Step 3: Inference Phase\\
    \For{GPU $i=0$ to $N-1$ \textbf{in parallel}}{
        \For{$m^{(i)}=0$ to $M_{\text{warmup}}-1$}{
            Conduct one-step inference on $p_i$ with buffer.\\
            Update buffer synchronously.\\
            Get $\tilde{x}_{t_{m}}$ by synchronous \texttt{All-Gather}
        }
        \For{$m^{(i)}=M_{\text{warmup}}$ to $M_{i}$}{
            Conduct one-step inference on $p_i$ with buffer.\\
            \If{GPU $i$ is fast}{
                \If{$(m^{(i)}-M_{\text{warmup}})\%2==0$}{
                    Update buffer asynchronously.\\
                    Get $\tilde{x}_{t_{m}}$ by synchronous \texttt{All-Gather}
                }
                \Else{
                    No update for buffer.\\
                    Get $\tilde{x}_{t_{m}}$ from last step.
                }
            }
            \Else{
                Update buffer by asynchronously.\\
                Get $\tilde{x}_{t_{m}}$ by synchronous \texttt{All-Gather}
            }
        }
    }
    \textbf{Return:} Denoised image $\tilde{x}_{t_M}$
\end{algorithm}

\subsection{System Model and Problem Formulation}
\label{subsec:system_model}

We consider a system comprising \(N\) heterogeneous GPUs, each with a relative computational capability \(c_i > 0\) (the fastest GPU is normalized to \(c_i = 1\), and slower devices have \(c_i < 1\)) and a background utilization \(\rho_i \in [0, 1]\) (0 idle, 1 fully loaded).
Parameter \(c_i\) can be obtained offline via benchmarking, while \(\rho_i\) is measured via system APIs to capture current workload.

Building on the DDIM sampling process, our method generates images from an initial noise through \(M\) denoising steps, where \(M\) governs a trade-off between generation fidelity and computation latency. To accelerate this process, we adopt the patch parallelism strategy in DistriFusion by partitioning the initial noise into spatial patches processed concurrently across multiple GPUs.

Given an initial noise \(x_{t_0}\) and a pre-trained diffusion model $\epsilon_\theta(\cdot,\cdot)$, the problem is to find an optimal scheduling strategy, i.e., a set of assignments $(M_i, P_i)$ for each GPU $i$ that minimizes the end-to-end generation latency $R_{\text{total}}$. 
Specifically, $R_{\text{total}}$ is governed by the per-GPU processing latency $R_i(M_i, P_i, c_i, \rho_i)$, which is the computation time for GPU $i$ to complete $M_i$ steps on patch $p_i$ under current load $\rho_i$.

Our proposed strategy, \name, addresses this optimization problem by creating a joint spatio-temporal workload distribution. Unlike prior methods like DistriFusion that employ a uniform temporal assignment, \name innovatively co-adjusts both $M_i$ and $P_i$.  
By tailoring both the spatial and temporal workload to each GPU's specific capabilities and current load, \name achieves a truly balanced execution.

\subsection{Temporal Adaptation: Flexible Steps Scheduling}
\label{subsec:temporal_adaptation}
To mitigate latency from straggler GPUs, our temporal adaptation mechanism -- referred to as the tiered step reduction strategy -- adaptively schedules the number of DDIM steps for each GPU. The process begins with a crucial warm-up phase, where all GPUs execute a uniform number of steps $M_{\text{warmup}}$ to ensure model convergence.
The core strategy begins with estimating the {effective speed} $v_i$ for each GPU based on $c_i$ and $\rho_i$.
We identify the fastest GPU with speed $v_{\max}$ and assign it a base step number $M_{\text{base}}$. We then assign each $M_i$ according to the following formulation: 
\begin{equation}
    M_i = 
        \begin{cases} 
        M_{\text{base}} & \text{if } a\cdot v_{\max} < v_i \leq v_{\max}, \\ 
        \frac{1}{2}M_{\text{base}}+\frac{1}{2}M_{\text{warmup}} & \text{if } b\cdot v_{\max} < v_i \leq a\cdot v_{\max}, \\
        \text{(Excluded)}& \text{if } v_i \leq b\cdot v_{\max}, 
        \end{cases}
    \label{eq:steps_coordinating}
\end{equation}
where $0<b<a<1$ are the hyperparameters, representing a trade-off between latency and quality. The thresholds $a$ and $b$ are designed to capture the hardware heterogeneity inherent in multi-GPU systems. A higher $a$ forces a larger proportion of GPUs to perform step reduction, which minimizes latency but introduces a potential compromise in image fidelity due to the coarser denoising process. Conversely, a smaller $a$ shifts the load-balancing burden toward Spatial Adaptation, preserving image quality at the cost of potential straggler-induced delays. $b$ defines the system's operational boundary. A high $b$ ensures high-efficiency execution by excluding excessively slower nodes. However, an overly aggressive $b$ may lead to resource underutilization. Due to the diversity of hardware combinations, detailed parameter tuning is beyond the scope of this paper. Nevertheless, the qualitative guidelines we provide are sufficient to ensure stable operation of the system in most heterogeneous environments.

\begin{figure}[tp]
    \centering
    \includegraphics[width=0.98\linewidth]{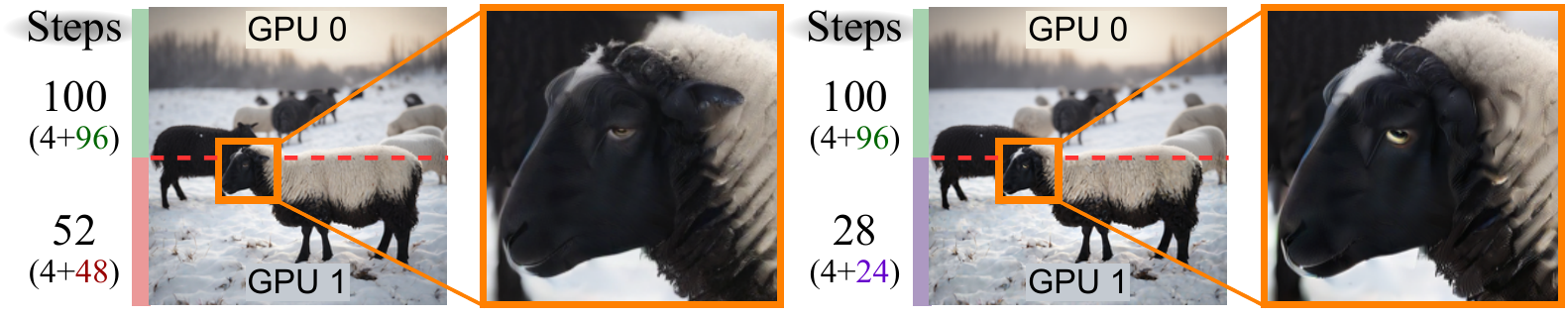}
    \caption{
    Detailed image texture comparison between standard (left) and deeper (right) step reduction strategy. Zoomed-in boxes highlight the artifacts and loss of semantic consistency caused by excessive step pruning on slower GPU.
    }
    \label{fig:insight1}
\end{figure}

A key design choice in Eq.~\eqref{eq:steps_coordinating} is the reduction ratio for straggler GPUs. While more aggressive reduction could further diminish latency, it risks catastrophic quality loss. To justify our choice of a 1/2 reduction, we investigate a \textit{deeper step reduction} strategy, which reduces post-warm-up steps on stragglers to 1/4. As illustrated in Figure~\ref{fig:insight1}, the qualitative comparison reveals that a significantly coarser denoising trajectory on slower GPUs leads to a pronounced degradation of semantic details and fine textures. These artifacts are clearly visible in the zoomed-in regions, where anatomical inconsistencies (e.g., missing ears) and blurred pupil details are prominent when steps are overly reduced.

Consequently, our strategy of pruning steps for slower GPUs (the 1/2 coefficient) represents a more balanced equilibrium for practical deployment, which significantly reduces the overall per-interval synchronization bottleneck without compromising generative fidelity. Furthermore, to avoid severe performance degradation, the extremely slow GPUs ($v_i \leq bv_{\max}$) are excluded. This adaptive scheduling effectively balances the computation load, minimizing per-interval bottlenecks and significantly reducing overall latency.

\subsection{Spatial Adaptation: Fine-grained Patch Size Mending}
\label{subsec:spatial_adaptation}

\begin{figure}[tp]
    \centering
    \includegraphics[width=0.98\linewidth]{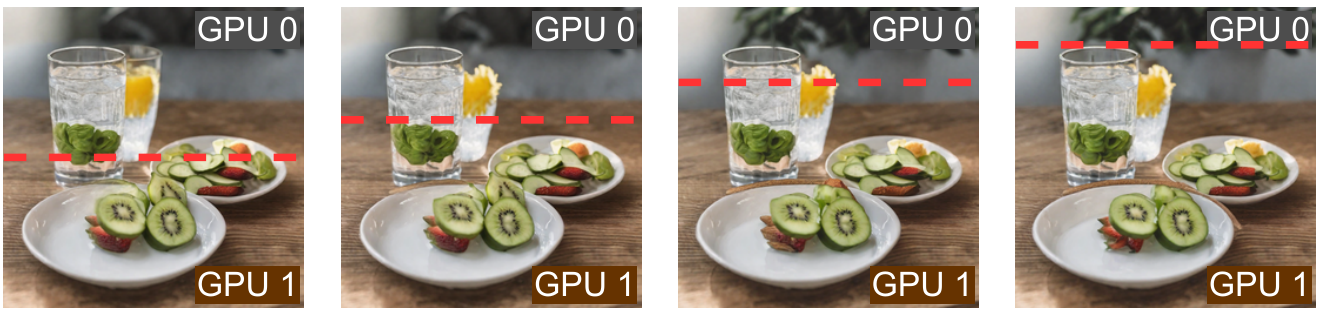}
    \caption{
    Quality visualization under varying patch assignment ratios. The red dashed line indicates the adaptive split boundary determined by Patch Size Mending. Consistent perceptual quality confirms that \name effectively decouples workload balancing from image synthesis without introducing tiling artifacts.
    }
    \label{fig:insight2}
\end{figure}

Temporal Adaptation alone leaves a residual workload imbalance, as all GPUs still process patches of uniform size. Our spatial adaptation mechanism, {\em Patch Size Mending}, resolves this by allocating patch sizes $P_i$ proportional to each GPU's effective processing rate, defined as $\frac{v_i}{M_i}$ to account for both its speed  $v_i$ and assigned steps $M_i$.

To achieve global load balance, we equalize the effective computational rate $\frac{v_i/M_i}{P_i}$ across all GPUs, subject to the spatial coverage constraint \(\sum P_i = 1\). This yields the optimal patch size for each GPU i:
\[
\frac{v_0/M_0}{P_0} = \frac{v_1/M_1}{P_1} = \dots = \frac{v_{N-1}/M_{N-1}}{P_{N-1}},
\]

\begin{equation}
    P_i = \frac{v_i/M_i}{\sum_{j=0}^{N-1} v_j/M_j} \cdot P_{\text{total}},
    \label{eq:patchmending}
\end{equation}
where $P_{\text{total}}$ is the spatial size of the full image. This formulation mends the residual imbalance from Temporal Adaptation. Consequently, all GPUs finish processing within similar latency per synchronization interval. 

Crucially, while Patch Size Mending flexibly redistributes the workload, it does not compromise generative performance. As visualized in Figure~\ref{fig:insight2}, the image quality remains consistent across various patch assignment ratios. This confirms that our spatial redistribution is seamless, effectively decoupling workload balancing from the synthesis process while preserving the structural integrity and fidelity of the generated content. In practice, $P_{\text{total}}$ must also satisfy hardware/operator constraints (e.g., power-of-two dimensions) -- these implementation details are explained in our experimental configuration.

While a straightforward design might rely solely on SA, our evaluations reveal that purely spatial partitioning yields diminishing returns as hardware disparity grows. This is primarily because latency scaling deviates from linearity when fixed system overheads become dominant. Moreover, excessive partitioning granularity can lead to GPU underutilization and instability. To overcome this scaling limit, \name introduces TA, which mitigates bottlenecks at a coarser level by reducing the total computational load on slower GPUs. By synergizing TA’s workload reduction with SA’s fine-grained calibration, the system maximizes hardware utilization, achieving speedups significantly superior to spatial redistribution alone while preserving high generative fidelity.

In summary, \name achieves a twofold latency reduction: 1) Temporal Adaptation frees up more computing resources on slower GPUs and reduces the coarse-grained workload at the same time, avoiding them becoming bottlenecks. 2) Spatial Adaptation then provides fine-grained calibration by reallocating image patches, thereby maximizing the collective utilization of all available computational resources.

\section{Theoretical Analysis}
\subsection{Temporal Redundancy in DistriFusion}
\label{redundancy-analysis}
The concept of {\em temporal redundancy} was first proposed by PipeFusion\cite{pipefusion}, who formalized the phenomenon observed in DistriFusion\cite{distrifusion} -- the small gap between inputs and activations across consecutive diffusion denoising steps, which can be defined as $\left|\Delta\tilde{x}_{t_m}\right|=\left| \tilde{x}_{t_{m}} - \tilde{x}_{t_{m+1}} \right|$. Such temporal redundancy enables the potential for parallelizing inference. 

In this subsection, we will quantitatively analyze the temporal redundancy present in first-order solvers.

\begin{thm}[Asymptotic bound of temporal redundancy in DistriFusion]
\label{thm:temporalredundancy}
    Under the assumptions of bounded states, model outputs, and the smooth noise schedule, the discretization error of the latent variable $\tilde{x}_t$ between adjacent steps satisfies the asymptotic bound:
    \begin{equation}
    \label{eq:thm1}
        \left| \tilde{x}_{t_{m}} - \tilde{x}_{t_{m+1}} \right| = \mathcal{O}\left(\frac{1}{M}\right).
    \end{equation}
    We denote $\tilde{x}_{t_m}^{(i)}$ as the output of the $i$-th GPU at step $t_m$. Since multiple GPUs share identical inference steps, the results after each synchronization are semantically aligned. This yields $\left|\tilde{x}_{t_{m}}^{(i)} - \tilde{x}_{t_{m+1}}^{(j)} \right| =\left| \tilde{x}_{t_{m}} - \tilde{x}_{t_{m+1}} \right|$ for any devices $i$ and $j$, hence Equation~\eqref{eq:thm1} represents the asymptotic bound of temporal redundancy in DistriFusion.
\end{thm}

\begin{proof}[proof sketch]
The proof relies on the Taylor expansion of the DDIM update equation with respect to the step size $h$. Assuming the noise schedule $\alpha$ is continuously differentiable and bounded, we expand the coefficients in the update rule $\Delta \tilde{x}_m = \tilde{x}_{t_{m+1}} - \tilde{x}_{t_m}$. 
Specifically, the ratio $\alpha_{t_{m+1}}/\alpha_{t_m}$ and the exponential term $e^{h_{m+1}}$ are linearized around $h=0$. This yields a relationship of the form:
\begin{equation}
    \tilde{x}_{t_{m+1}} - \tilde{x}_{t_{m}} = h_{m+1} \cdot \Phi + \mathcal{O}(h_{m+1}^2),
\end{equation}
where $\Phi$ is a bounded term dependent on the model output and schedule derivatives. Since the step size satisfies $h_{m+1} = \Theta(1/M)$, the per-step difference is dominated by the linear term, concluding that $|\Delta \tilde{x}_m| = \mathcal{O}(1/M)$. 
\end{proof}
\para{Remark.} By assuming a non-zero velocity field ($\|\Phi\| > 0$) and applying the reverse triangle inequality, we can further show that the lower bound is $\Omega(1/M)$, strengthening the result to a tight asymptotic bound of $\Theta(1/M)$.


Theorem~\ref{thm:temporalredundancy} establishes that the magnitude of the step-wise difference $\left|\Delta\tilde{x}_{t_m}\right|$ is strictly of the order $\mathcal{O}(1/M)$. Empirical experiments by DistriFusion demonstrate that exploiting this temporal redundancy -- characterized by the $\mathcal{O}(1/M)$ scaling--does not compromise output quality. This property guarantees convergence in patch parallelism and is effectively leveraged to accelerate inference in multi-GPU systems.

\subsection{Temporal Redundancy in \name}
The preceding section established that temporal redundancy in DistriFusion is bounded by $\mathcal{O}(1/M)$, allowing computation-communication overlap across GPUs. 
We will now extend this theorem to analyze the temporal redundancy of \name when the inference steps on each GPU are not consistent.

\begin{thm} [Asymptotic bound of temporal redundancy in \name]
\label{thm:orchestra}
    Under the assumptions of bounded states, model outputs, and the smooth noise schedule, for two devices $i$ and $j$ whose step count satisfies $nM_i=M_j=M$, the discretization error of their latent variables satisfies:
    \begin{equation*}
        \left|\tilde{x}_{t_{m}^i}^{(i)}-\tilde{x}_{t_{m+n}^j}^{(j)}\right|=\mathcal{O}\left(\frac{1}{M}\right),
    \end{equation*}
    where $t_m^i=t_m^j=s$ and $t_{m+1}^i=t_{m+n}^j=t$.
\end{thm}

\begin{proof}[proof sketch]
To bound the difference between unaligned latents across two devices, we introduce an intermediate state: the latent on device $i$ at the step $t_{m+1}^i$, which aligns with the step $t_{m+n}^j$. This decomposition isolates two sources of error: the trajectory discrepancy between the two devices at this aligned step, and the temporal redundancy on device $i$ between its current and next step. Based on the error analysis in DPM-Solver, the aligned trajectory discrepancy is bounded by $\mathcal{O}(n^2/M^2)$. Meanwhile, the single-device temporal redundancy is bounded by $\mathcal{O}(1/M)$ as established in Theorem~\ref{thm:temporalredundancy}. Finally, by applying the triangle inequality, the total error is the sum of these components. Since the first-order temporal term $\mathcal{O}(1/M)$ dominates the second-order trajectory error, the total asynchronous redundancy follows the $\mathcal{O}(1/M)$ bound.
\end{proof}

Theorem~\ref{thm:orchestra} characterizes the asymptotic bound of temporal redundancy using different step schedulers, with an order consistent with using homogeneous step schedulers in Theorem~\ref{thm:temporalredundancy}.
This implies a significant capability: {\em Even when step schedulers operate with dissimilar steps, we can still orchestrate the inference across multiple GPUs} with appropriate communication operations to align activation information. 

\section{Evaluation}
\label{sec:experiment}

This section presents the experimental evaluation of our proposed method. We begin by detailing the experimental settings, followed by a comprehensive analysis of \name across various heterogeneous configurations. We then conduct extensive ablation studies, sensitivity analyses, and scalability tests to further validate its robustness. 

\subsection{Experimental Settings}
\label{ssec:exp_settings}

\begin{figure}[t]
    \centering
    \includegraphics[width=0.98\linewidth]{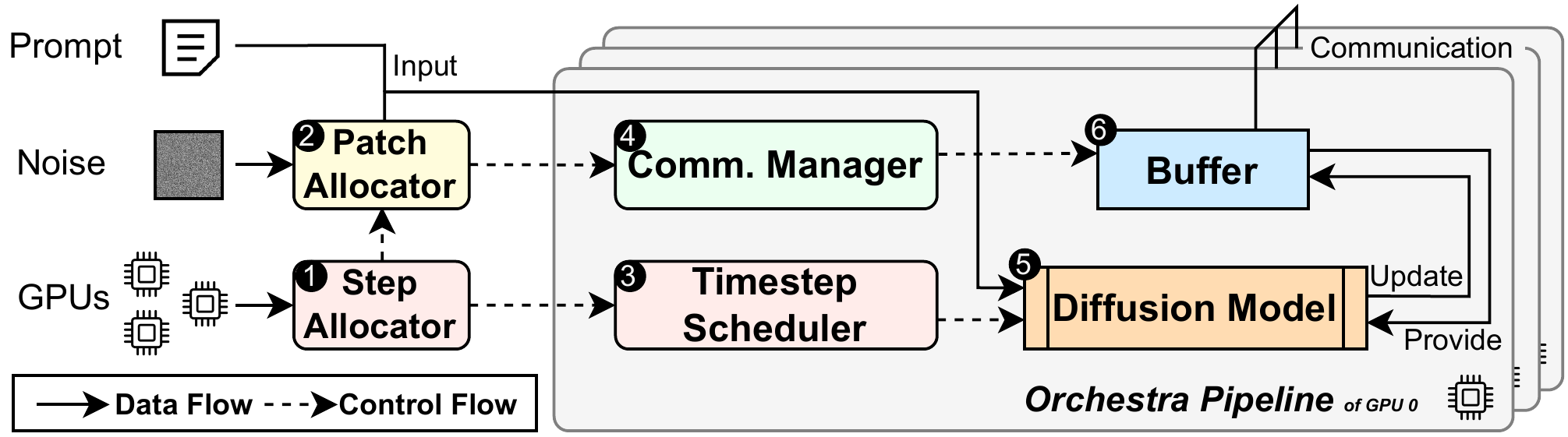}
    \caption{System architecture.}
    \label{fig:system}
\end{figure}

\para{Evaluation Setup.}
Our experiments utilize two GPU nodes with distinct configurations tailored to different computational scales. Node A is equipped with two consumer-grade NVIDIA GeForce RTX 4090 GPUs interconnected via PCIe, while Node B comprises eight data-center-grade NVIDIA A100 GPUs interconnected via NVLink.
To simulate realistic heterogeneous computing capacities and background workload interference, we deploy a specialized occupancy program on the target GPU before inference. The program adjusts tensor size to stabilize utilization at a preset level, after which the tensor size remains fixed. Consequently, the ``occupancy'' reported in our experiments serves as a quantitative proxy for  heterogeneous capabilities, allowing us to evaluate \name in environments with varying available compute capacities.


\begin{figure}[t]
    \centering
    \begin{subfigure}[c]{0.49\linewidth}
        \centering
        \includegraphics[width=\linewidth]{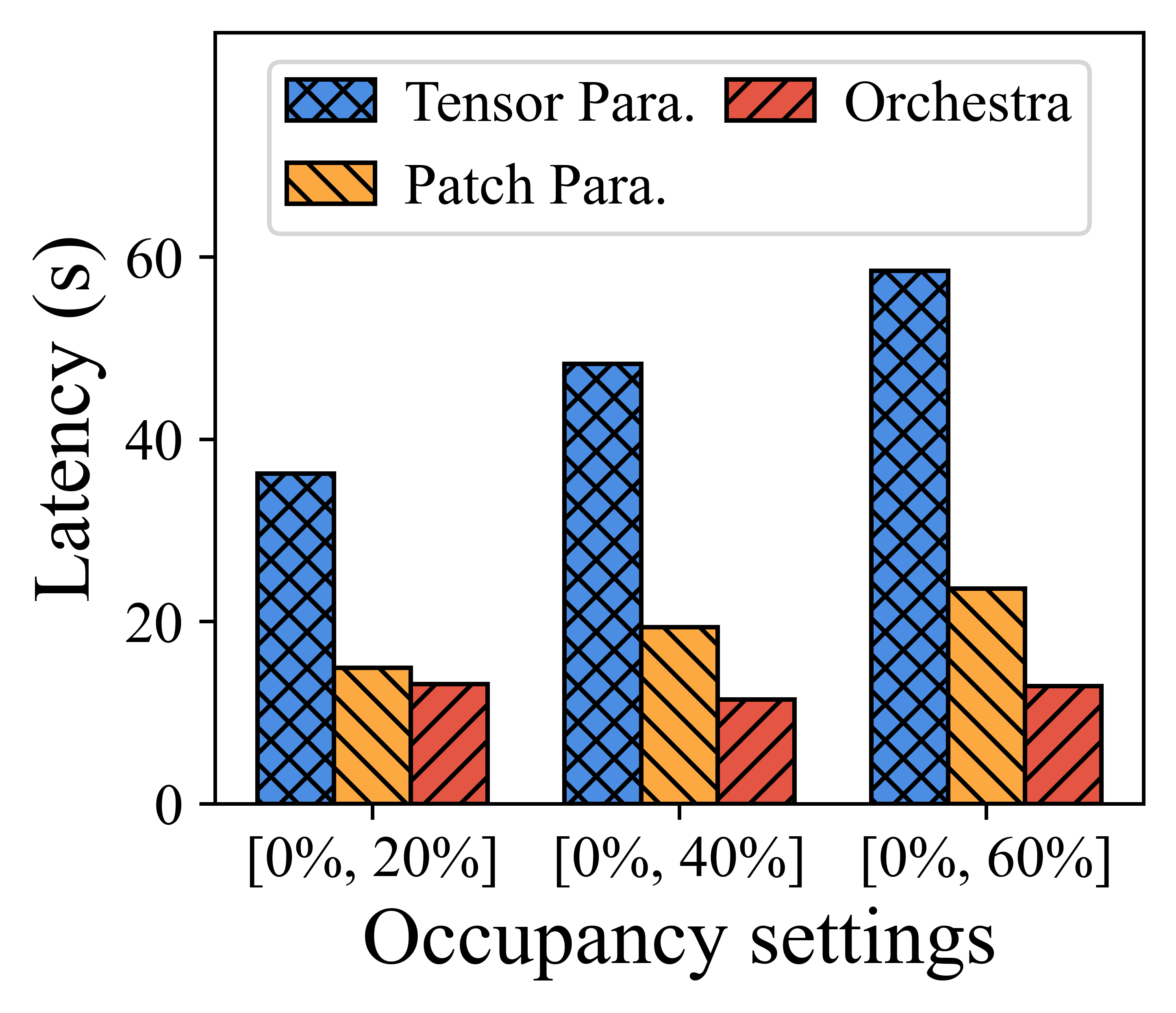}
        \caption{Increasing total occupancy.}
        \label{fig:latency_A}
    \end{subfigure}
    \hfill 
    \begin{subfigure}[c]{0.49\linewidth}
        \centering
        \includegraphics[width=\linewidth]{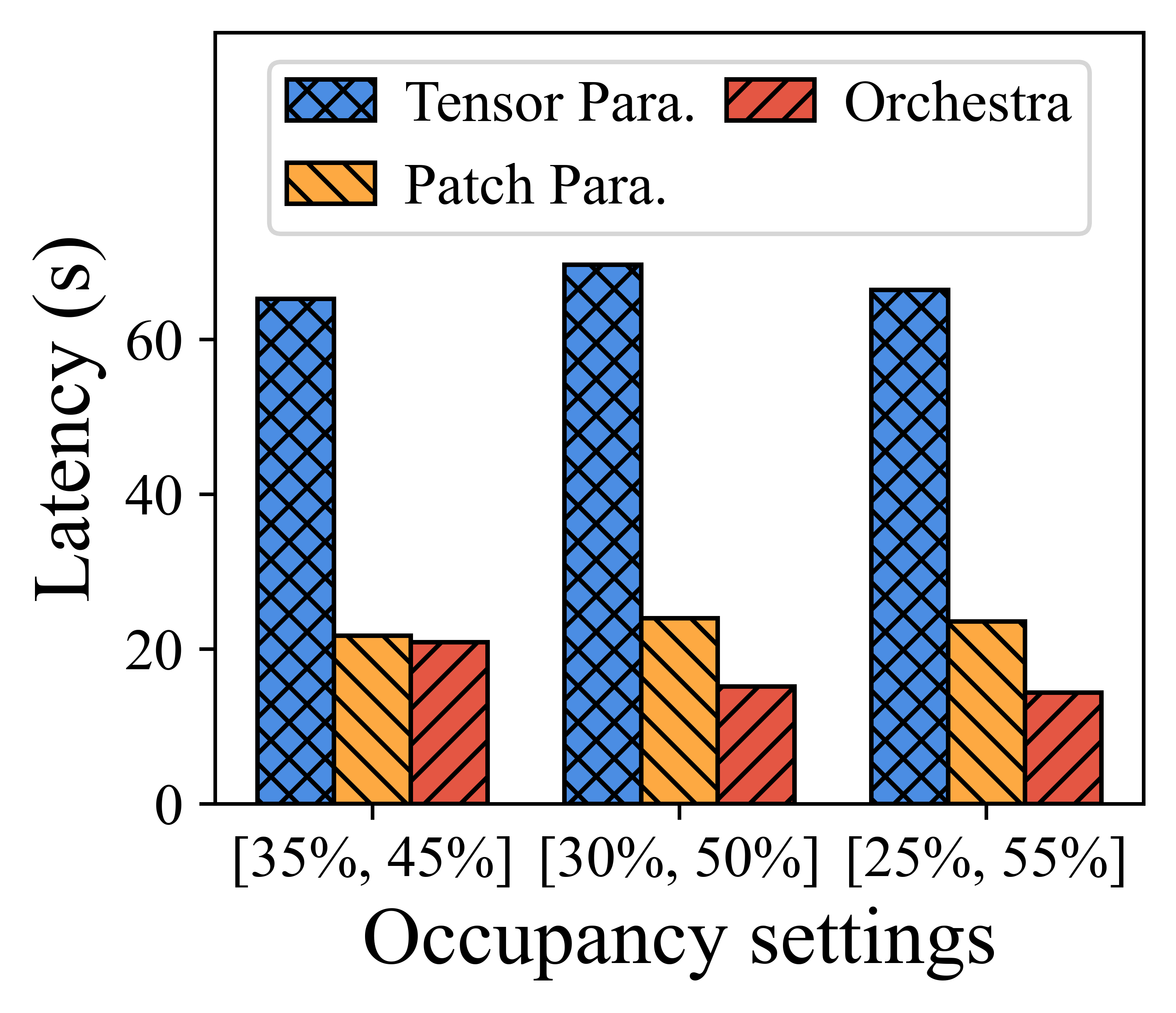}
        \caption{Fixed total occupancy.}
        \label{fig:latency_B}
    \end{subfigure}

    \caption{Latency comparison across various occupancy settings.}
    \label{fig:latency}
\end{figure}


\begin{table}[tp]
    \centering
    \resizebox{0.45\textwidth}{!}{
        \begin{tabular}{@{}c c c c c@{}}
        \toprule
        \multirow{2}{*}{\textbf{Method}} & \textbf{Occupancy} & \textbf{Patch} & \multicolumn{2}{c}{\textbf{FID} ($\downarrow$)} \\
        \cmidrule{4-5}
        & \textbf{Settings} & \textbf{Assignment} & w/G.T. & w/Orig. \\
        \midrule
        Patch Para.            & -          & \textcolor{lhgreen}{16}:\textcolor{lhgreen}{16} & 17.8 & 0.9~ \\
        \midrule
        \multirow{6}{*}{\name} & 0\%, 20\%  & \textcolor{lhgreen}{18}:\textcolor{lhgreen}{14} & 17.8 & 1.0~ \\
                               & 0\%, 40\%  & \textcolor{lhgreen}{15}:\textcolor{lhred}{17}   & 18.0 & 1.1~ \\
                               & 0\%, 60\%  & \textcolor{lhgreen}{16}:\textcolor{lhred}{16}   & 18.0 & 1.4~ \\
        \cmidrule{2-5}
                               & 35\%, 45\% & \textcolor{lhgreen}{17}:\textcolor{lhgreen}{15} & 17.7 & 1.0~ \\
                               & 30\%, 50\% & \textcolor{lhgreen}{13}:\textcolor{lhred}{19}   & 18.0 & 1.3~ \\
                               & 25\%, 55\% & \textcolor{lhgreen}{14}:\textcolor{lhred}{18}   & 17.9 & 1.1~ \\
        \bottomrule
        \end{tabular}
    }
    \caption{Quality comparison (FID) under various occupancy settings and its corresponding patch assignments. Patch size in \textcolor{lhred}{red}/\textcolor{lhgreen}{green} means the steps are reduced/not reduced.}
    
    \label{tab:quality2}
\end{table}
\begin{figure*}[t]
    \centering
    \includegraphics[width=0.995\textwidth]{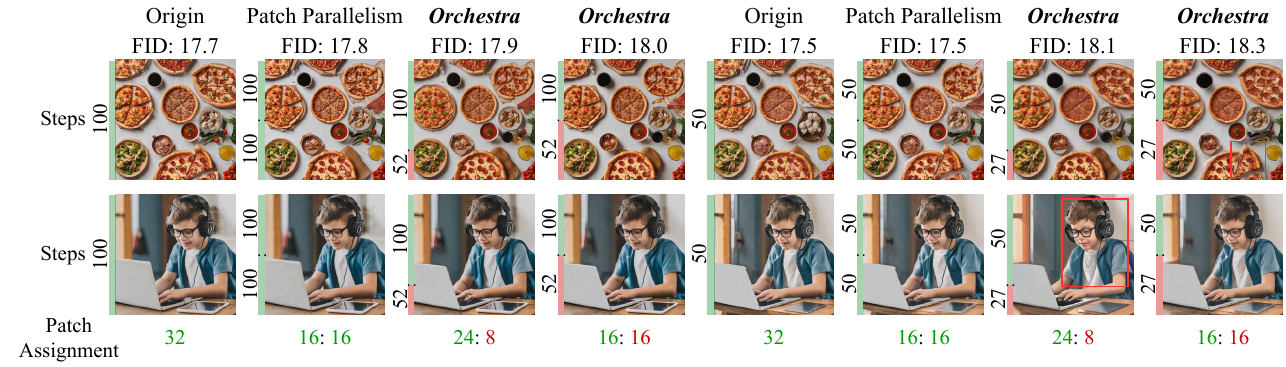}
    \caption{
    Image quality visualization across various patch size for step reduction. The FID values are compared with ground truth images. Patch size in \textcolor{lhred}{red} (\textcolor{lhgreen}{green}) means the steps are (not) reduced.
    }
    \label{fig:quality}
\end{figure*}

\begin{table*}[t]
\centering
    \resizebox{0.95\textwidth}{!}{
        \begin{tabular}{cccc*{2}{c}*{2}{c}*{2}{c}}
        \toprule
        \multirow{2}{*}{$M_\text{base}$} & \multirow{2}{*}{\textbf{Method}} & \multirow{2}{*}{\textbf{Patch Assignment}} & \multirow{2}{*}{\textbf{Steps}} & 
        \multicolumn{2}{c}{\textbf{PSNR} ($\uparrow$)} & 
        \multicolumn{2}{c}{\textbf{LPIPS} ($\downarrow$)} & 
        \multicolumn{2}{c}{\textbf{FID} ($\downarrow$)} \\
        \cmidrule(lr){5-6} \cmidrule(lr){7-8} \cmidrule(lr){9-10}
         & & & & w/ G.T. & w/ Orig. & w/ G.T. & w/ Orig. & w/ G.T. & w/ Orig. \\
        \midrule
        \multirow{5}{*}{100} & Origin & -- & 100 & 9.52 & -- & 0.794 & -- & \textbf{17.7} & -- \\
        & Patch Parallelism & \textcolor{lhgreen}{16}:\textcolor{lhgreen}{16} & 100:100 & 9.52 & \textbf{24.65} & 0.794 & \textbf{0.143} & 17.8 & \textbf{0.9} \\
        \cmidrule{2-10}
        & \multirow{3}{*}{\name} & \textcolor{lhgreen}{24}:\textcolor{lhred}{8} & 100:52  & \textbf{9.56} & 23.04 & 0.795 & 0.188 & 17.9 & 1.2 \\
        &  & \textcolor{lhgreen}{16}:\textcolor{lhred}{16} & 100:52 & 9.52 & 22.14 & 0.795 & 0.210 & 18.0 & 1.4 \\
        &  & \textcolor{lhgreen}{8}:\textcolor{lhred}{24} & 100:52 & 9.51 & 23.04 & \textbf{0.793} & 0.188 & 18.2 & 1.3 \\
        \midrule
        \addlinespace
        \multirow{5}{*}{50} & Origin & --& 50  & 9.64 & -- & 0.796 & -- & \textbf{17.5} & -- \\
        & Patch Parallelism & \textcolor{lhgreen}{16}:\textcolor{lhgreen}{16} & 50:50 & 9.62 & \textbf{24.67} & 0.797 & \textbf{0.145} & \textbf{17.5} & \textbf{0.9} \\
        \cmidrule{2-10}
        & \multirow{3}{*}{\name} & \textcolor{lhgreen}{24}:\textcolor{lhred}{8} & 50:27 & \textbf{9.66} & 23.32 & 0.801 & 0.202 & 18.1 & 1.4 \\
        &  & \textcolor{lhgreen}{16}:\textcolor{lhred}{16} & 50:27 & 9.62 & 22.32 & 0.799 & 0.234 & 18.3 & 1.9 \\
        &  & \textcolor{lhgreen}{8}:\textcolor{lhred}{24} & 50:27  & 9.61 & 23.13 & \textbf{0.795} & 0.217 & 18.2 & 2.0 \\
        \bottomrule
        \end{tabular}
    }
    \caption{
    Quality comparison. $\uparrow$/$\downarrow$: Higher/Lower is better. Patch size in \textcolor{lhred}{red}/\textcolor{lhgreen}{green} means the steps are reduced/not reduced. Results are reported for $M_{\text{base}} \in \{50, 100\}$ across multiple metrics. ``w/ G.T.'' and ``w/ Orig.'' denote metrics relative to ground truth and original outputs, respectively.
    }
    \label{tab:quality1}
\end{table*}
\para{Diffusion Model.} 
We employ Stable Diffusion XL (SDXL) \cite{SDXL}, a 2.3B-parameter model, for two primary reasons: (1) its state-of-the-art, high-resolution image generation quality is representative of demanding real-world applications; and (2) its hybrid architecture, which integrates convolutional and attention layers, is characteristic of modern large-scale diffusion models.

\para{Baselines.} 
Comparative evaluation includes two state-of-the-art distributed inference paradigms: 
patch parallelism from DistriFusion~\cite{distrifusion} and tensor parallelism. Tensor parallelism enables distributed diffusion inference by employing synchronous all-reduce operations at each computational layer. For fairness, all baselines were re-implemented using identical NCCL communication primitives.

\para{Evaluation Metrics.} 
Performance is quantified through end-to-end latency and speedup ratio. Image quality is assessed using Peak Signal-to-Noise Ratio (PSNR), LPIPS, and Fréchet Inception Distance (FID) on the COCO Captions 2014 validation set \cite{COCO}. We use partial captions to generate images. PSNR measures pixel-level accuracy between two images. LPIPS measures the perceptual similarity. FID compares sets of generated and real images using a pretrained model to evaluate overall quality and semantic similarity.


\para{System Architecture.} Figure~\ref{fig:system} illustrates the system architecture of the proposed \name pipeline system. The diagram delineates the processing sequence from inputs (prompt and noise) through core modules (e.g., patch allocator, timestep scheduler, communication manager) to output generation via the diffusion model. The arrows explicitly trace the data flow and control flow executed on one GPU.

\para{Implementation Details.}
To ensure a fair comparison, no other acceleration techniques (e.g., batch splitting for CFG, CUDA graphs) are applied to the patch parallelism.
Unless in experiments involving image quality considerations or otherwise specified, we set $M_\text{base}=100$, $M_\text{warmup}=4$. To satisfy computational constraints of certain operators' inputs, we set $P_\text{total}=32$. The parameters $a$ and $b$ control the thresholds for temporal adaptation and GPU usage, respectively. Unless otherwise specified, we set these values to $a=0.75$ and $b=0.25$.
The effective speed $v_i$ for each GPU is derived directly from historical inference time profiles. All reported values are averaged over five runs.

\para{All-Gather for uneven-sized tensors.}
It is worth mentioning that uneven patch sizes require customized communication. Early versions of \texttt{torch} lacked an all-gather implementation adapted for uneven sized tensors. While \texttt{all\_gather\_object} inherently requires a synchronous communication to keep data types and sizes consistent from other devices, it cannot satisfy our asynchronous communication requirements. To address it, we developed two asynchronous communication approaches: one achieves this by padding tensors to uniform sizes before performing \texttt{all\_gather}, while the other uses multiple asynchronous \texttt{broadcast} to emulate \texttt{all\_gather}. Both approaches can still mask communication latency within computation.

\subsection{Main Results}
\subsubsection{Latency.}
We benchmark \name on Node A for $1024 \times 1024$ image generation under two scenarios: (1) Increasing Total Occupancy ([0\%, 20\%], [0\%, 40\%], [0\%, 60\%]) representing decreasing aggregate resources; and (2) Fixed Total Occupancy ([35\%, 45\%], [30\%, 50\%], [25\%, 55\%]), maintaining total load while increasing resource  asymmetry.

As shown in Figure~\ref{fig:latency}, \name consistently outperforms  baselines across diverse occupancy settings. Tensor parallelism is the slowest in every setting and is severely affected by workload occupancy. In the first scenario in Figure~\ref{fig:latency}(a), \name achieves latency reductions of 12\% to 45\% over the patch parallelism. When total occupancy is fixed at 80\% but redistributed in Figure~\ref{fig:latency}(b), \name achieves 4\% to 39\% latency reductions. Notably, when GPU computational capabilities are different yet comparable, the specific setting of $a$ in Equation~\ref{eq:steps_coordinating} does not trigger the Temporal Adaptation. For example, in the [0\%, 20\%] and [35\%, 45\%] cases, load balancing is achieved solely by adjusting the patch size, which leads to suboptimal speedup.

\subsubsection{Quality.} 
In our algorithm, the image quality is mainly influenced by $M_\text{base}$ and the patch applied steps reduction. We first compare the performance of \name with other baselines across various patch sizes for step reduction. 

The quantitative results in Table~\ref{tab:quality1} and qualitative visualizations in Figure~\ref{fig:quality} collectively demonstrate the impact of step reduction on generation quality. 
As shown in Table~\ref{tab:quality1}, the {\it{Origin}} method performs non-distributed inference, and the images it generates are called the original images, while the ground truth images refer to COCO validation set. Patch parallelism generally achieves higher PSNR values with original images when compared to \name, as it does not reduce the steps on any patch. \name tends to have slightly lower PSNR with original images but higher values with ground truth images, and moreover, the difference between the FID calculated from ground truth images and other methods is minimal (less than 1). 
As illustrated in Figure~\ref{fig:quality}, although reducing $M_\text{base}$ and applying step reduction (highlighted by red boxes) introduce minor pixel-level variations, the core semantic information remains consistent with the original images. This aligns with the intuition that sampling iterations influence visual refinement. However, the discrepancies in \name are marginal and do not undermine the overall content integrity.

Furthermore, Table~\ref{tab:quality2} evaluates the quantitative results of \name under various occupancy settings (consistent with the settings in Figure~\ref{fig:latency}). Results indicate that \name can select the appropriate inference steps and patch size for different GPUs, the FID scores remain remarkably stable. The FID relative to ground truth fluctuates minimally between 17.7 and 18.0, while the FID relative to the original images stays within a low range (1.0 to 1.4). These results confirm that \name effectively balances computational efficiency with high-quality generative performance.

\begin{table}[tp]
    \centering
    \resizebox{0.48\textwidth}{!}{
        \begin{tabular}{@{}c l l l l@{}}
        \toprule
        \multirow{2}{*}{\textbf{Occ.}} & \multicolumn{4}{c}{{\name}} \\
        \cmidrule(lr){2-5}
        & {None}
        & {+SA}
        & {+TA}
        & {+TA\&+SA} \\
        \midrule
        0\%, 20\% & 14.91s 
        & 13.16s \scriptsize\itshape 1.13×
        & 11.30s \scriptsize\itshape 1.32× 
        & \textbf{10.85s} \scriptsize\itshape \textbf{1.37}× \\
        \midrule
        0\%, 40\% & 19.41s 
        & 15.13s \scriptsize\itshape 1.28× 
        & 12.05s \scriptsize\itshape 1.61× 
        & \textbf{11.43s} \scriptsize\itshape \textbf{1.70}× \\
        \midrule
        0\%, 60\% & 23.67s 
        & 17.73s \scriptsize\itshape 1.34× 
        & 13.03s \scriptsize\itshape 1.82× 
        & \textbf{12.92s} \scriptsize\itshape \textbf{1.83}× \\
        \bottomrule
        \end{tabular}
    }
    \caption{Ablation study. ``+SA'' means Spatial Adaptation while ``+TA'' means Temporal Adaptation.}
    
    \label{tab:ablation}
\end{table}

\subsection{Ablation Study}

\para{The impact of patch size.}
To isolate and evaluate the efficacy of exclusively utilizing Spatial Adaptation (SA) for load balancing under heterogeneous GPU utilization, we disable Temporal Adaptation (TA) to maintain uniform steps across all GPUs. In this configuration, load balancing is achieved solely by SA based on their available computational capacity. 

\begin{figure}[tp]
    \centering
    \includegraphics[width=0.98\linewidth]{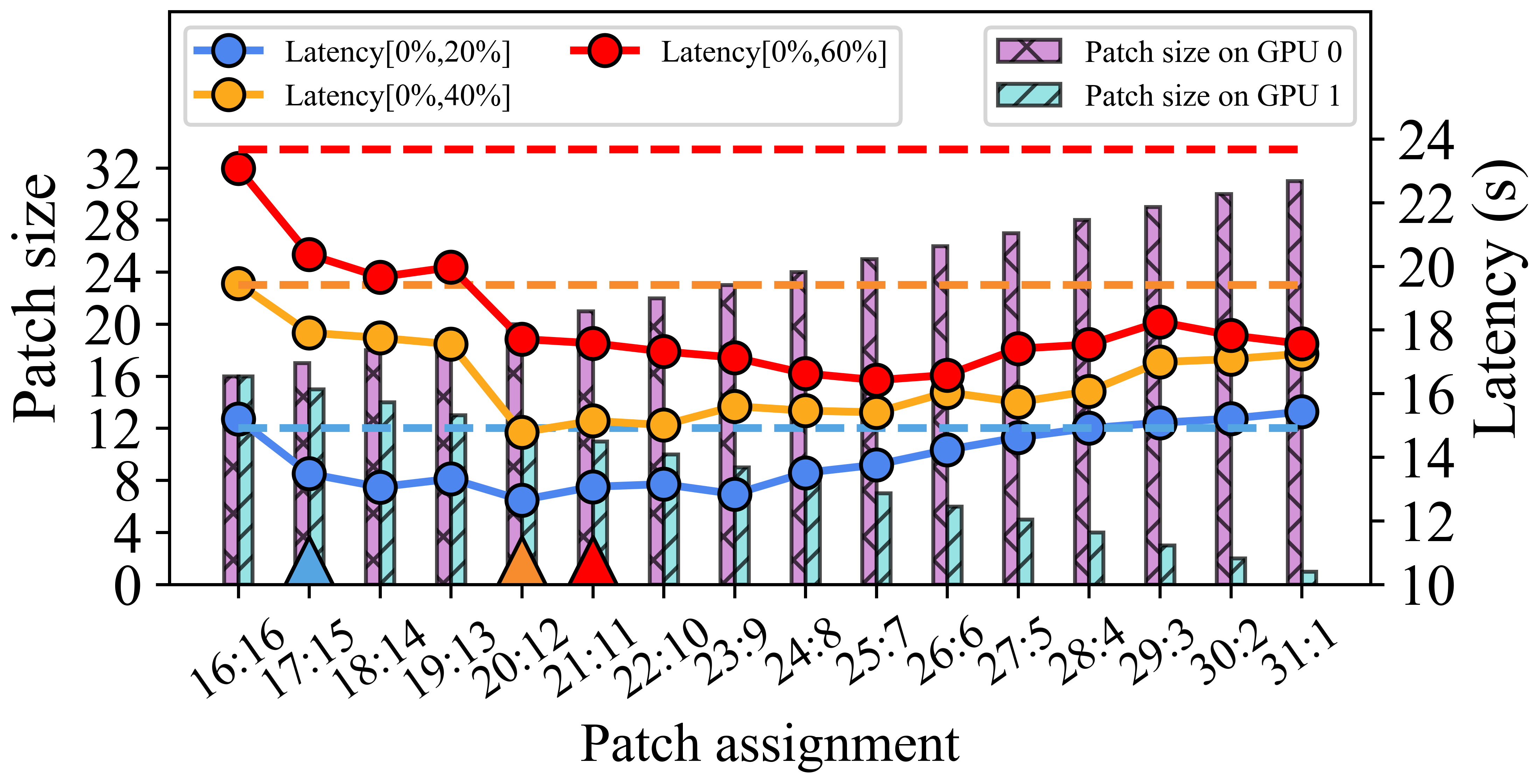}
    \caption{
    The inference latency under different occupancies and patch sizes. Different colors represent distinct occupancy settings. Dashed lines in corresponding colors indicate pure patch parallelism's latency under each configuration, while triangles mark the actual patch assignment selected by \name. 
    }
    \label{fig:spatial_manually}
\end{figure}

As shown in Table~\ref{tab:ablation}, SA alone improves inference speed by 1.12$\times$--1.34$\times$ by mitigating the straggler effect caused by the slowest GPU.
However, we observe diminishing returns in extreme cases. As illustrated in Figure~\ref{fig:spatial_manually}, when the load disparity is excessively large, patch allocation guided by perceived computational capacity may deviate from the optimum. This is because the single-step latency ceases to maintain a strictly linear relationship with patch size due to inherent fixed overheads.

\para{The impact of steps reduction.} TA directly addresses the synchronization bottleneck by adaptively reducing post-warm-up inference steps on slower GPUs. As evidenced in Table~\ref{tab:ablation}, the standard strategy of halving post-warm-up steps liberates substantial computational resources on constrained devices, achieving up to 1.82$\times$ speedup under [0\%, 60\%] occupancy. This demonstrates TA’s superior efficacy in mitigating latency induced by severely heterogeneous GPUs. 

\subsection{Sensitivity Analysis}
\para{warm-up steps.} 
We conducted a sensitivity analysis on the number of warm-up steps, specifically focusing on the trade-off between inference acceleration and generative quality. For these experiments, base diffusion steps are fixed at 100 under [0\%, 60\%] occupancy. 
As illustrated in Table~\ref{tab:warmup}, \name consistently outperforms patch parallelism across all $M_{\text{warmup}}$ values. Notably, the speedup diminishes as $M_{\text{warmup}}$ increases. This is because a larger $M_{\text{warmup}}$ constrains the optimization window for load-balancing, forcing the system to remain straggler-bound by the slowest GPU for a longer duration.

\begin{figure}[tp]
    \centering
    \includegraphics[width=0.98\linewidth]{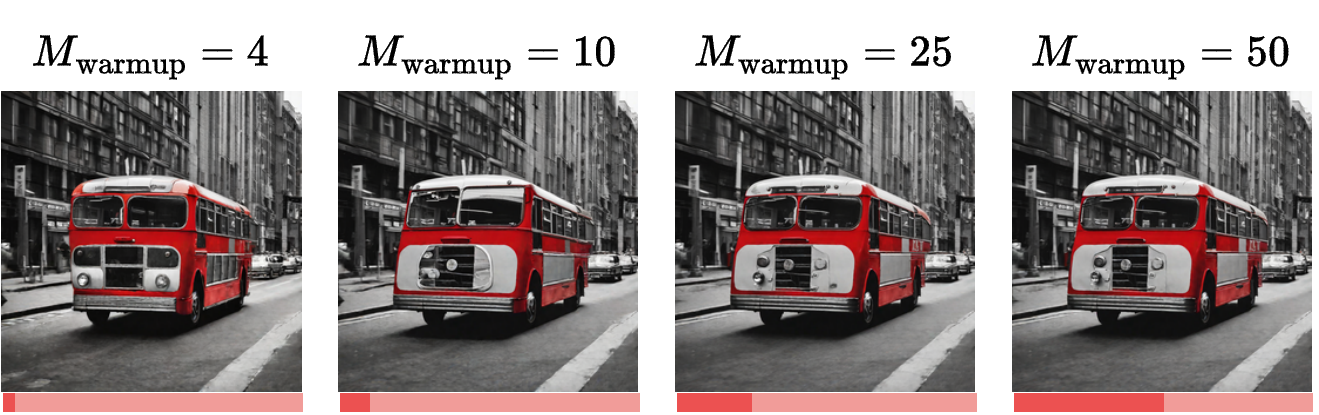}
    \caption{Image quality visualization across various warm-up steps. We fixed the total number of steps $M_\text{base}=100$.}
    \label{fig:quality_warmup}
\end{figure}

Crucially, while one might expect a longer warm-up phase to enhance generative performance, our qualitative analysis in Figure~\ref{fig:quality_warmup} reveals that increasing $M_{\text{warmup}}$ yields limited improvement in final image quality. 
Given that excessive warm-up steps incur a significant latency penalty without providing substantial quality gains, we conclude that a shorter, well-calibrated warm-up phase strikes the most effective balance for practical deployment.

\begin{table}[tp]
    \centering
    \resizebox{0.45\textwidth}{!}{
        \begin{tabular}{@{}c l l l l l@{}}
        \toprule
        $M_{\text{warmup}}$ & 2 & 4 & 10 & 25 & 50 \\
        \midrule
        Patch Para. & 23.24s & 23.44s & 23.86s & 24.36s & 25.75s~ \\
        \midrule
        \name & 13.19s & 13.56s & 14.15s & 16.52s & 20.30s \\
        \midrule
        \textbf{Speedup} & \textbf{1.76×} & \textbf{1.73×} & \textbf{1.69×} & \textbf{1.47×} & \textbf{1.27×} \\
        \bottomrule
        \end{tabular}
    }
    \caption{Latency comparison with various warm-up steps. We fixed the total number of steps $M_\text{base}=100$.}
    
    \label{tab:warmup}
\end{table}

\subsection{Scalability analysis.}

\begin{figure}[t]
    \centering
    \includegraphics[width=0.98\linewidth]{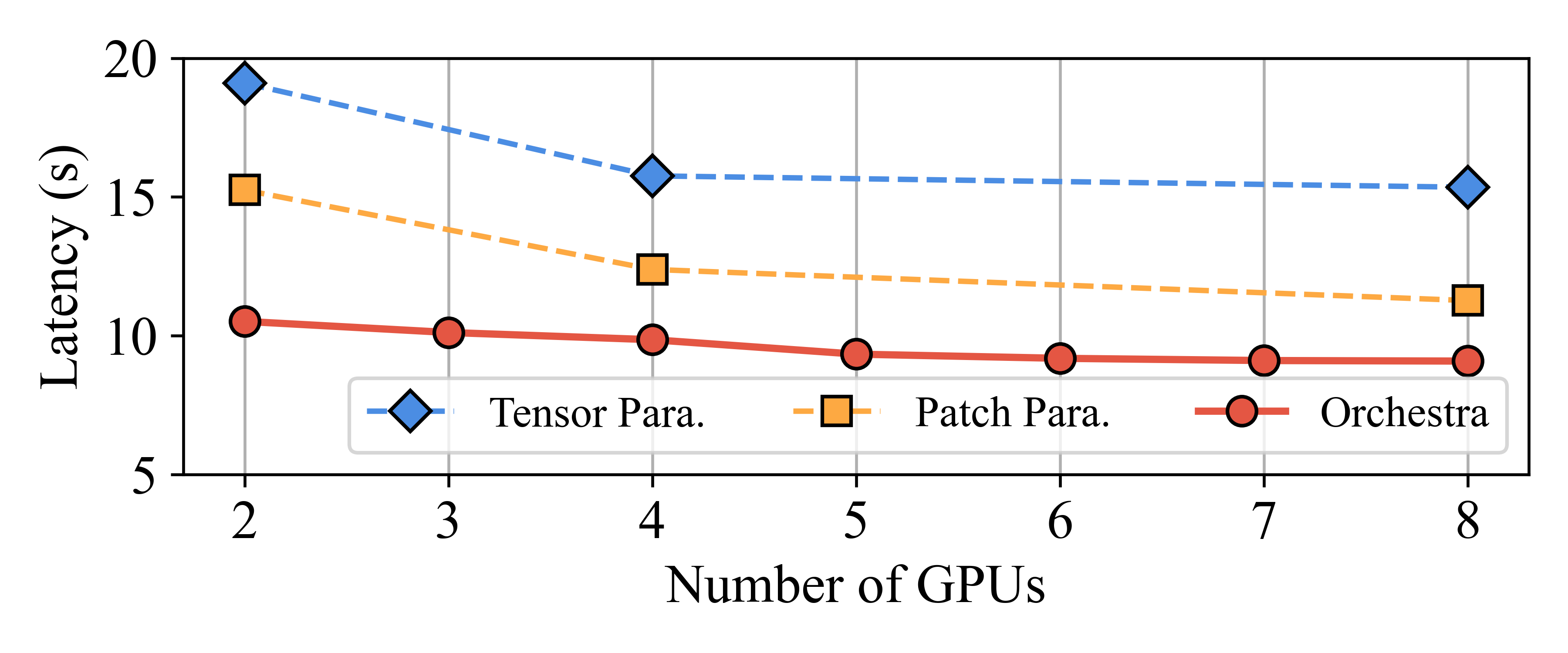}
    \caption{Scalability comparison. We evaluate the latency of \name and other baselines across various GPU counts.}
    \label{fig:scalability}
\end{figure}

To evaluate the scalability of \name, we conducted experiments on Node B, which is equipped with eight NVIDIA A100 GPUs interconnected via NVLink. To simulate a heterogeneous environment, we designated one GPU as a straggler with a 70\% occupancy. We evaluated the end-to-end inference latency as the total GPU count scaled from 2 to 8.

As shown in the Fig.~\ref{fig:scalability}, \name consistently achieves the lowest latency across all GPU counts.
Unlike tensor parallelism, which is bottlenecked by frequent synchronous overhead, or patch parallelism, which remains vulnerable to straggler-induced delays, \name maintains superior efficiency through its adaptive scheduling. 
Notably, while these conventional frameworks typically restrict partitioning to power-of-two configurations (e.g., 2, 4, 8), \name's fine-grained adaptation enables seamless execution on arbitrary GPUs counts (e.g., 3, 5, 6, 7), which offers significantly higher scheduling flexibility. 



While maintaining the lowest latency, \name exhibits diminishing returns at higher GPU counts. This is consistent with our expectations, as the current prototype does not yet incorporate CUDA Graph to mitigate kernel launch overheads, nor does it employ batch splitting techniques specifically optimized for CFG. While integrating these orthogonal optimizations would likely yield substantial further acceleration, a detailed exploration is beyond the scope of this paper due to space constraints. Nonetheless, \name remains highly effective in real-world environments by adaptively calculating $P_i$ and $M_i$ to accommodate irregular resource availability, ensuring a balanced workload distribution that maximizes overall cluster utility.
\section{Conclusion}

In this work, we presented \name, a novel framework that tackles the latency bottleneck of diffusion model inference on heterogeneous multi-GPU systems. By uniquely exploiting the iterative nature of diffusion sampling, \name introduces fine-grained DDIM step scheduling to free up computational resources on slower GPUs and employ patch size mending for spatial load balancing. This dual-axis strategy achieves significant latency reductions for SDXL inference without compromising image quality, demonstrating its effectiveness on prevalent heterogeneous hardware. 
Future work will focus on experiments with large-scale clusters, exploring synergies with techniques like model compression, extending to harder tasks (e.g., video generation), and further optimizations.

\clearpage
\bibliographystyle{IEEEtran}
\bibliography{main.bib}

@article{DDPM,
  title={Denoising diffusion probabilistic models},
  author={Ho, Jonathan and Jain, Ajay and Abbeel, Pieter},
  journal={Advances in neural information processing systems},
  volume={33},
  pages={6840--6851},
  year={2020}
}

@article{DDIM,
  title={Denoising diffusion implicit models},
  author={Song, Jiaming and Meng, Chenlin and Ermon, Stefano},
  journal={arXiv preprint arXiv:2010.02502},
  year={2020}
}

@article{DPMsolver,
  title={Dpm-solver: A fast ode solver for diffusion probabilistic model sampling in around 10 steps},
  author={Lu, Cheng and Zhou, Yuhao and Bao, Fan and Chen, Jianfei and Li, Chongxuan and Zhu, Jun},
  journal={Advances in Neural Information Processing Systems},
  volume={35},
  pages={5775--5787},
  year={2022}
}

@article{DPMsolverplus,
  title={Dpm-solver++: Fast solver for guided sampling of diffusion probabilistic models},
  author={Lu, Cheng and Zhou, Yuhao and Bao, Fan and Chen, Jianfei and Li, Chongxuan and Zhu, Jun},
  journal={arXiv preprint arXiv:2211.01095},
  year={2022}
}

@article{song2020score, 
  title={Score-based generative modeling through stochastic differential equations},
  author={Song, Yang and Sohl-Dickstein, Jascha and Kingma, Diederik P and Kumar, Abhishek and Ermon, Stefano and Poole, Ben},
  journal={arXiv preprint arXiv:2011.13456},
  year={2020}
}

@inproceedings{SD,
  title={High-resolution image synthesis with latent diffusion models},
  author={Rombach, Robin and Blattmann, Andreas and Lorenz, Dominik and Esser, Patrick and Ommer, Bj{\"o}rn},
  booktitle={Proceedings of the IEEE/CVF conference on computer vision and pattern recognition},
  pages={10684--10695},
  year={2022}
}

@inproceedings{DiT,
  title={Scalable diffusion models with transformers},
  author={Peebles, William and Xie, Saining},
  booktitle={Proceedings of the IEEE/CVF international conference on computer vision},
  pages={4195--4205},
  year={2023}
}

@article{SDXL,
  title={Sdxl: Improving latent diffusion models for high-resolution image synthesis},
  author={Podell, Dustin and English, Zion and Lacey, Kyle and Blattmann, Andreas and Dockhorn, Tim and M{\"u}ller, Jonas and Penna, Joe and Rombach, Robin},
  journal={arXiv preprint arXiv:2307.01952},
  year={2023}
}

@article{ho2022video,
  title={Video diffusion models},
  author={Ho, Jonathan and Salimans, Tim and Gritsenko, Alexey and Chan, William and Norouzi, Mohammad and Fleet, David J},
  journal={Advances in neural information processing systems},
  volume={35},
  pages={8633--8646},
  year={2022}
}

@article{ho2022imagen,
  title={Imagen video: High definition video generation with diffusion models},
  author={Ho, Jonathan and Chan, William and Saharia, Chitwan and Whang, Jay and Gao, Ruiqi and Gritsenko, Alexey and Kingma, Diederik P and Poole, Ben and Norouzi, Mohammad and Fleet, David J and others},
  journal={arXiv preprint arXiv:2210.02303},
  year={2022}
}

@article{blattmann2023stable,
  title={Stable video diffusion: Scaling latent video diffusion models to large datasets},
  author={Blattmann, Andreas and Dockhorn, Tim and Kulal, Sumith and Mendelevitch, Daniel and Kilian, Maciej and Lorenz, Dominik and Levi, Yam and English, Zion and Voleti, Vikram and Letts, Adam and others},
  journal={arXiv preprint arXiv:2311.15127},
  year={2023}
}

@inproceedings{popov2021grad,
  title={Grad-tts: A diffusion probabilistic model for text-to-speech},
  author={Popov, Vadim and Vovk, Ivan and Gogoryan, Vladimir and Sadekova, Tasnima and Kudinov, Mikhail},
  booktitle={International conference on machine learning},
  pages={8599--8608},
  year={2021},
  organization={PMLR}
}

@article{shen2023naturalspeech,
  title={Naturalspeech 2: Latent diffusion models are natural and zero-shot speech and singing synthesizers},
  author={Shen, Kai and Ju, Zeqian and Tan, Xu and Liu, Yanqing and Leng, Yichong and He, Lei and Qin, Tao and Zhao, Sheng and Bian, Jiang},
  journal={arXiv preprint arXiv:2304.09116},
  year={2023}
}

@inproceedings{MM-Diffusion,
  title={Mm-diffusion: Learning multi-modal diffusion models for joint audio and video generation},
  author={Ruan, Ludan and Ma, Yiyang and Yang, Huan and He, Huiguo and Liu, Bei and Fu, Jianlong and Yuan, Nicholas Jing and Jin, Qin and Guo, Baining},
  booktitle={Proceedings of the IEEE/CVF Conference on Computer Vision and Pattern Recognition},
  pages={10219--10228},
  year={2023}
}

@article{CoDi,
  title={Any-to-any generation via composable diffusion},
  author={Tang, Zineng and Yang, Ziyi and Zhu, Chenguang and Zeng, Michael and Bansal, Mohit},
  journal={Advances in Neural Information Processing Systems},
  volume={36},
  pages={16083--16099},
  year={2023}
}

@article{kingma2021variational,
  title={Variational diffusion models},
  author={Kingma, Diederik and Salimans, Tim and Poole, Ben and Ho, Jonathan},
  journal={Advances in neural information processing systems},
  volume={34},
  pages={21696--21707},
  year={2021}
}

@article{xue2023sa,
  title={Sa-solver: Stochastic adams solver for fast sampling of diffusion models},
  author={Xue, Shuchen and Yi, Mingyang and Luo, Weijian and Zhang, Shifeng and Sun, Jiacheng and Li, Zhenguo and Ma, Zhi-Ming},
  journal={Advances in Neural Information Processing Systems},
  volume={36},
  pages={77632--77674},
  year={2023}
}

@article{bao2022analytic,
  title={Analytic-dpm: an analytic estimate of the optimal reverse variance in diffusion probabilistic models},
  author={Bao, Fan and Li, Chongxuan and Zhu, Jun and Zhang, Bo},
  journal={arXiv preprint arXiv:2201.06503},
  year={2022}
}

@article{pipefusion,
  title={Pipefusion: Displaced patch pipeline parallelism for inference of diffusion transformer models},
  author={Wang, Jiannan and Fang, Jiarui and Li, Aoyu and Yang, PengCheng},
  journal={arXiv e-prints},
  pages={arXiv--2405},
  year={2024}
}

@inproceedings{distrifusion,
  title={Distrifusion: Distributed parallel inference for high-resolution diffusion models},
  author={Li, Muyang and Cai, Tianle and Cao, Jiaxin and Zhang, Qinsheng and Cai, Han and Bai, Junjie and Jia, Yangqing and Li, Kai and Han, Song},
  booktitle={Proceedings of the IEEE/CVF Conference on Computer Vision and Pattern Recognition},
  pages={7183--7193},
  year={2024}
}

@article{asyncdiff,
  title={Asyncdiff: Parallelizing diffusion models by asynchronous denoising},
  author={Chen, Zigeng and Ma, Xinyin and Fang, Gongfan and Tan, Zhenxiong and Wang, Xinchao},
  journal={arXiv preprint arXiv:2406.06911},
  year={2024}
}

@article{shih2023parallel,
  title={Parallel sampling of diffusion models},
  author={Shih, Andy and Belkhale, Suneel and Ermon, Stefano and Sadigh, Dorsa and Anari, Nima},
  journal={Advances in Neural Information Processing Systems},
  volume={36},
  pages={4263--4276},
  year={2023}
}

@article{xDiT,
  title={xDiT: an Inference Engine for Diffusion Transformers (DiTs) with Massive Parallelism},
  author={Fang, Jiarui and Pan, Jinzhe and Sun, Xibo and Li, Aoyu and Wang, Jiannan},
  journal={arXiv preprint arXiv:2411.01738},
  year={2024}
}

@article{parastep,
  title={Communication-Efficient Diffusion Denoising Parallelization via Reuse-then-Predict Mechanism},
  author={Wang, Kunyun and Li, Bohan and Yu, Kai and Guo, Minyi and Zhao, Jieru},
  journal={arXiv e-prints},
  pages={arXiv--2505},
  year={2025}
}

@inproceedings{Clipper,
  title={Clipper: A $\{$Low-Latency$\}$ online prediction serving system},
  author={Crankshaw, Daniel and Wang, Xin and Zhou, Guilio and Franklin, Michael J and Gonzalez, Joseph E and Stoica, Ion},
  booktitle={14th USENIX Symposium on Networked Systems Design and Implementation (NSDI 17)},
  pages={613--627},
  year={2017}
}

@inproceedings{Nexus,
  title={Nexus: A GPU cluster engine for accelerating DNN-based video analysis},
  author={Shen, Haichen and Chen, Lequn and Jin, Yuchen and Zhao, Liangyu and Kong, Bingyu and Philipose, Matthai and Krishnamurthy, Arvind and Sundaram, Ravi},
  booktitle={Proceedings of the 27th ACM Symposium on Operating Systems Principles},
  pages={322--337},
  year={2019}
}

@article{zhang2024mixtran,
  title={Mixtran: an efficient and fair scheduler for mixed deep learning workloads in heterogeneous GPU environments},
  author={Zhang, Xiao},
  journal={Cluster Computing},
  volume={27},
  number={3},
  pages={2775--2784},
  year={2024},
  publisher={Springer}
}

@inproceedings{zhang2024hap,
  title={Hap: Spmd dnn training on heterogeneous gpu clusters with automated program synthesis},
  author={Zhang, Shiwei and Diao, Lansong and Wu, Chuan and Cao, Zongyan and Wang, Siyu and Lin, Wei},
  booktitle={Proceedings of the Nineteenth European Conference on Computer Systems},
  pages={524--541},
  year={2024}
}

@inproceedings{choi2022serving,
  title={Serving heterogeneous machine learning models on $\{$Multi-GPU$\}$ servers with $\{$Spatio-Temporal$\}$ sharing},
  author={Choi, Seungbeom and Lee, Sunho and Kim, Yeonjae and Park, Jongse and Kwon, Youngjin and Huh, Jaehyuk},
  booktitle={2022 USENIX Annual Technical Conference (USENIX ATC 22)},
  pages={199--216},
  year={2022}
}

@inproceedings{kong2025ppipe,
  title={$\{$PPipe$\}$: Efficient Video Analytics Serving on Heterogeneous $\{$GPU$\}$ Clusters via $\{$Pool-Based$\}$ Pipeline Parallelism},
  author={Kong, Z Jonny and Xu, Qiang and Hu, Y Charlie},
  booktitle={2025 USENIX Annual Technical Conference (USENIX ATC 25)},
  pages={679--698},
  year={2025}
}

@article{shoeybi2019megatron,
  title={Megatron-lm: Training multi-billion parameter language models using model parallelism},
  author={Shoeybi, Mohammad and Patwary, Mostofa and Puri, Raul and LeGresley, Patrick and Casper, Jared and Catanzaro, Bryan},
  journal={arXiv preprint arXiv:1909.08053},
  year={2019}
}

@article{jacobs2023deepspeed,
  title={Deepspeed ulysses: System optimizations for enabling training of extreme long sequence transformer models},
  author={Jacobs, Sam Ade and Tanaka, Masahiro and Zhang, Chengming and Zhang, Minjia and Song, Shuaiwen Leon and Rajbhandari, Samyam and He, Yuxiong},
  journal={arXiv preprint arXiv:2309.14509},
  year={2023}
}

@article{fang2024usp,
  title={Usp: A unified sequence parallelism approach for long context generative ai},
  author={Fang, Jiarui and Zhao, Shangchun},
  journal={arXiv preprint arXiv:2405.07719},
  year={2024}
}

@inproceedings{mohammed2020distributed,
  title={Distributed inference acceleration with adaptive DNN partitioning and offloading},
  author={Mohammed, Thaha and Joe-Wong, Carlee and Babbar, Rohit and Di Francesco, Mario},
  booktitle={IEEE INFOCOM 2020-IEEE conference on computer communications},
  pages={854--863},
  year={2020},
  organization={IEEE}
}

@inproceedings{hu2022distributed,
  title={Distributed inference with deep learning models across heterogeneous edge devices},
  author={Hu, Chenghao and Li, Baochun},
  booktitle={IEEE INFOCOM 2022-IEEE Conference on Computer Communications},
  pages={330--339},
  year={2022},
  organization={IEEE}
}

@article{zhao2024hetegen,
  title={Hetegen: Heterogeneous parallel inference for large language models on resource-constrained devices},
  author={Zhao, Xuanlei and Jia, Bin and Zhou, Haotian and Liu, Ziming and Cheng, Shenggan and You, Yang},
  journal={arXiv preprint arXiv:2403.01164},
  year={2024}
}

@article{tang2021aeml,
  title={AEML: An acceleration engine for multi-GPU load-balancing in distributed heterogeneous environment},
  author={Tang, Zhuo and Du, Lifan and Zhang, Xuedong and Yang, Li and Li, Kenli},
  journal={IEEE Transactions on Computers},
  volume={71},
  number={6},
  pages={1344--1357},
  year={2021},
  publisher={IEEE}
}

@article{COCO,
  title={Microsoft coco captions: Data collection and evaluation server},
  author={Chen, Xinlei and Fang, Hao and Lin, Tsung-Yi and Vedantam, Ramakrishna and Gupta, Saurabh and Doll{\'a}r, Piotr and Zitnick, C Lawrence},
  journal={arXiv preprint arXiv:1504.00325},
  year={2015}
}

@inproceedings{wupasta,
  title={PASTA: Training Acceleration for Vertical Federated Learning via Adaptive Pipeline Parallelism},
  author={Wu, Tian and Liu, Weijie and Liang, Han and Zhan, Ziwei and Duan, Jingpu and Wu, Chuan and Zuo, Jinhang and Chen, Xu and Zhang, Xiaoxi},
  booktitle={2025 IEEE/ACM 33rd International Symposium on Quality of Service (IWQoS)},
  year={2025},
  organization={IEEE}
}

@inproceedings{zhong2024poster,
  title={POSTER: In-network Model Inference for Distributed Systems via Programmable Switches},
  author={Zhong, Jianqiang and Li, Yihong and Liu, Shuo and Duan, Jingpu and Zhang, Xiaoxi and Chen, Xu},
  booktitle={Proceedings of the ACM SIGCOMM 2024 Conference: Posters and Demos},
  pages={75--77},
  year={2024}
}

@inproceedings{huang2024alleviating,
  title={Alleviating All-to-All Communication for Deep Learning Recommendation Model Inference},
  author={Huang, Songjun and Li, Yihong and Chen, Liangkun and Zhang, Xiaoxi and Liu, Shuo and Duan, Jingpu and Wu, Wenfei and Chen, Xu},
  booktitle={Workshop Proceedings of the 53rd International Conference on Parallel Processing},
  pages={104--105},
  year={2024}
}

\end{document}